\input harvmac.tex
\input epsf.tex



\lref\WittenYC{
E.~Witten,
``Phases of ${\cal N} = 2$ theories in two dimensions,''
Nucl.\ Phys.\ B {\bf 403}, 159 (1993);
{\tt arXiv:hep-th/9301042}.
}

\lref\ColemanUZ{
S.~R.~Coleman,
``More about the massive Schwinger model,''
Annals Phys.\  {\bf 101}, 239 (1976).
}

\lref\first{
J.~Maldacena and H.~Ooguri,
``Strings in $AdS_3$ and $SL(2,R)$ WZW model. Part 1: the spectrum,''
J.\ Math.\ Phys. 42 (2001) 2929
({\sl Special issue on M Theory}); {\tt arXiv/0001053}.
}

\lref\second{J.~Maldacena, H.~Ooguri and J.~Son,
``Strings in $AdS_3$ and the $SL(2,R)$ WZW model.
Part 2: Euclidean black hole,''
J.\ Math.\ Phys.  42 (2001) 2961
({\sl Special issue on M Theory}); {\tt
arXiv/0005183}.
}

\lref\siw{E.~Silverstein and E.~Witten,
``Criteria for conformal invariance of $(0,2)$ models,''
Nucl.\ Phys.\ B {\bf 444}, 161 (1995);
{\tt arXiv:hep-th/9503212}.
}

\lref\wittenjones{E.~Witten,
``Chern-Simons gauge theory as a string theory,''
{\tt arXiv:hep-th/9207094}.
}

\lref\wittenknot{E. Witten, ``Quantum field theory
and the Jones polynomial,'' Comm. Math. Phys. {\bf 121}, 351
(1989).}

\lref\bcov{M.~Bershadsky, S.~Cecotti, H.~Ooguri and C.~Vafa,
``Kodaira-Spencer theory of gravity and exact 
results for quantum string amplitudes,''
Commun.\ Math.\ Phys.\  {\bf 165}, 311 (1994);
{\tt arXiv:hep-th/9309140}.
}


{\Title{\vbox{
\hbox{CALT-68-2386, CITUSC/02-019}
\hbox{HUTP-02/A018}
\hbox{\tt hep-th/0205297}}}
{\vbox{
\centerline{Worldsheet Derivation of a Large $N$ Duality}}}
\vskip .3in
\centerline{Hirosi Ooguri$^1$ and Cumrun Vafa$^2$}
\vskip .4in
}

\centerline{$^1$ California Institute of Technology 452-48,
Pasadena, CA 91125, USA}
\vskip .1in
\centerline{$^2$ Jefferson Physical Laboratory, Harvard University,
Cambridge, MA 02138, USA}

\vskip .4in
We give a worldsheet proof of
the equivalence between the $U(N)$ Chern-Simons gauge 
theory on $S^3$ and the topological closed string theory on
the resolved conifold geometry. 
When the `t Hooft coupling
of the gauge theory is small, the dual closed string worldsheet
 develops a new branch.  We show that
the fluctuations of the worldsheet into this
branch effectively correspond to ``holes''
on the worldsheet, generating an open string sector. 
This leads to a microscopic description 
of how the `t Hooft expansion of gauge theory amplitudes
is reproduced in the closed string computation. 
We find that the closed string amplitudes also contain terms 
which are not captured in the `t Hooft expansion but are present 
in the exact computation in the gauge theory
 amplitudes.
These arise when the whole Riemann surface is in the
new branch. We also discuss the cases with $SO$ and $Sp$ gauge
groups.

\vfill
\eject

\newsec{Introduction}

Many examples of large $N$ dualities have been discovered
in string theory in recent years, in accord with `t Hooft's
original idea
\lref\tHooftJZ{
G.~'t Hooft,
``A planar diagram theory for strong interactions,''
Nucl.\ Phys.\ B {\bf 72}, 461 (1974).
} \tHooftJZ .  
Most of these results are motivated
from the viewpoint of target space physics.  However, the original
intuition of `t Hooft was based on worldsheet diagrams.  It is
thus natural to expect that one can derive large $N$ dualities
from the worldsheet perspective.  In 't Hooft's double line notation,
Feynman diagrams are expressed as surfaces with holes. 
It was postulated that somehow these holes get filled up, leading to 
closed Riemann surfaces without boundaries.
Given that we know what the closed string duals
are in many examples, it is natural to ask the reverse question.  
Namely, we could
try to see how the holes get developed from the closed string perspective,
leading to the Feynman diagrams of the gauge theories.
In this paper, we show how this works in the context of the duality
between the $U(N)$ Chern-Simons gauge theory on $S^3$ and the 
topological closed string theory
 on the resolved conifold, which was conjectured in 
\ref\gopva{R.~Gopakumar 
and C.~Vafa,
``On the gauge theory/geometry correspondence,''
Adv.\ Theor.\ Math.\ Phys.\  {\bf 3}, 1415 (1999);
{\tt arXiv:hep-th/9811131}.
 }.  By doing so, we also shed light on some aspects
of this duality and in particular how certain non-perturbative
terms in the gauge theory arise on the closed string dual.
We suspect that our derivation may well have
applications beyond the particular case considered here.

Let us first review the general setup of large $N$ dualities
proposed by `t Hooft.  Consider a gauge theory with $U(N)$ gauge 
group whose action is written as
$$S={1\over g_{{\rm YM}}^2} \int {\cal L}(A)$$
where $A$ is a gauge field.
A Feynman
diagram of $U(N)$ gauge theory drawn as a
``ribbon graph'' can be viewed as a closed 
Riemann surface with some holes. 
The dependence of the amplitude on the gauge coupling
constant $g_{{\rm YM}}^2$ and the rank $N$ of the gauge
group can be captured by the topology of this surface. 
For a ribbon graph with $g$ handles and $h$ holes, the amplitude
comes with the factor
$$(g_{{\rm YM}}^{2})^{-V+E}N^h=(g_{{\rm YM}}^2)^{-V+E-h}
(g_{{\rm YM}}^2 N)^h=(g_{{\rm YM}}^2)^{2g-2}
(g_{{\rm YM}}^2 N)^h=(g_{{\rm YM}}^2)^{2g-2}t^h, $$
where $V$ and $E$ are the number of vertices and propagators in
the graph, and we have defined $t=g_{{\rm YM}}^2 N$.
The full amplitude is given by summing over all topologies, 
\eqn\thooft{F=\sum_{g= 0}^\infty
\sum_{h=1}^\infty
\ (g_{{\rm YM}}^2)^{2g-2}t^h \ F_{g,h}=\sum_g 
(g_{{\rm YM}}^2)^{2g-2}\ F_g(t),}
where
\eqn\fg{F_g(t)=\sum_{h=1}^\infty \   t^h F_{g,h} }
and the coefficients $F_{g,h}$
may be a function of other parameters defining the theory.
The statement that this is
an expansion in powers of $1/N$ is the same as saying this is an
expansion in powers of $g_{{\rm YM}}^2$, as shown in the above, 
since $g_{{\rm YM}}^2 N$ is held fixed. 
The general conjecture of `t Hooft is that at large $N$,
with $g_{{\rm YM}}^2N=t$ held fixed, an equivalent description should
involve closed Riemann surfaces which are obtained from 
the ribbon graphs by  ``filling holes with disks.'' 
Compare Figure 1 of a ribbon graph and Figure 2 of the corresponding
Riemann surface:

\bigskip

\centerline{\epsfxsize 3truein \epsfysize 2truein\epsfbox{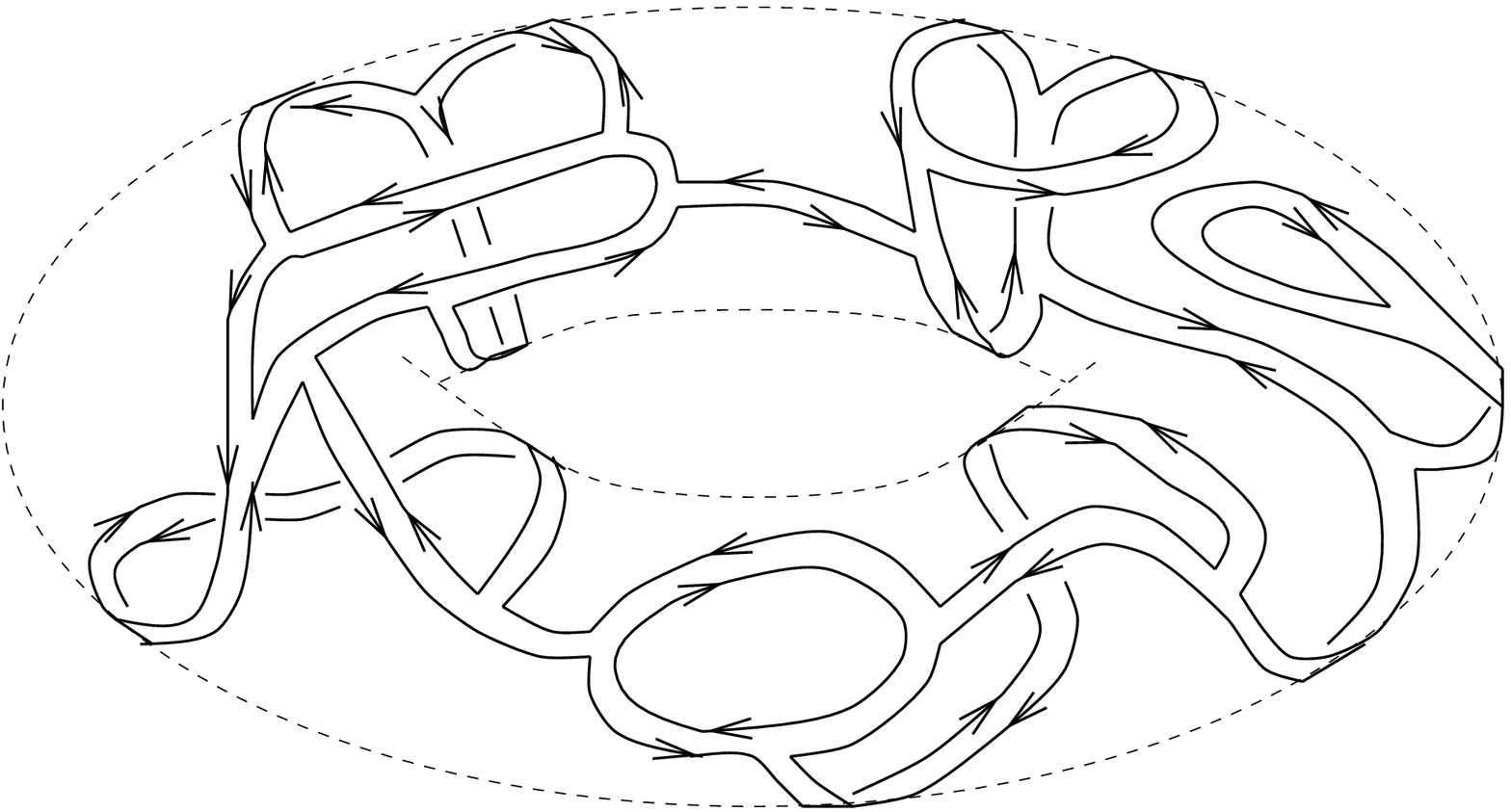}}
\noindent{\ninepoint\sl \baselineskip=8pt {\bf Figure 1}: {\sl
A Ribbon graph with $g=1$ and $h=9$.}}

\bigskip
\bigskip

\centerline{\epsfxsize 3truein \epsfysize 2truein\epsfbox{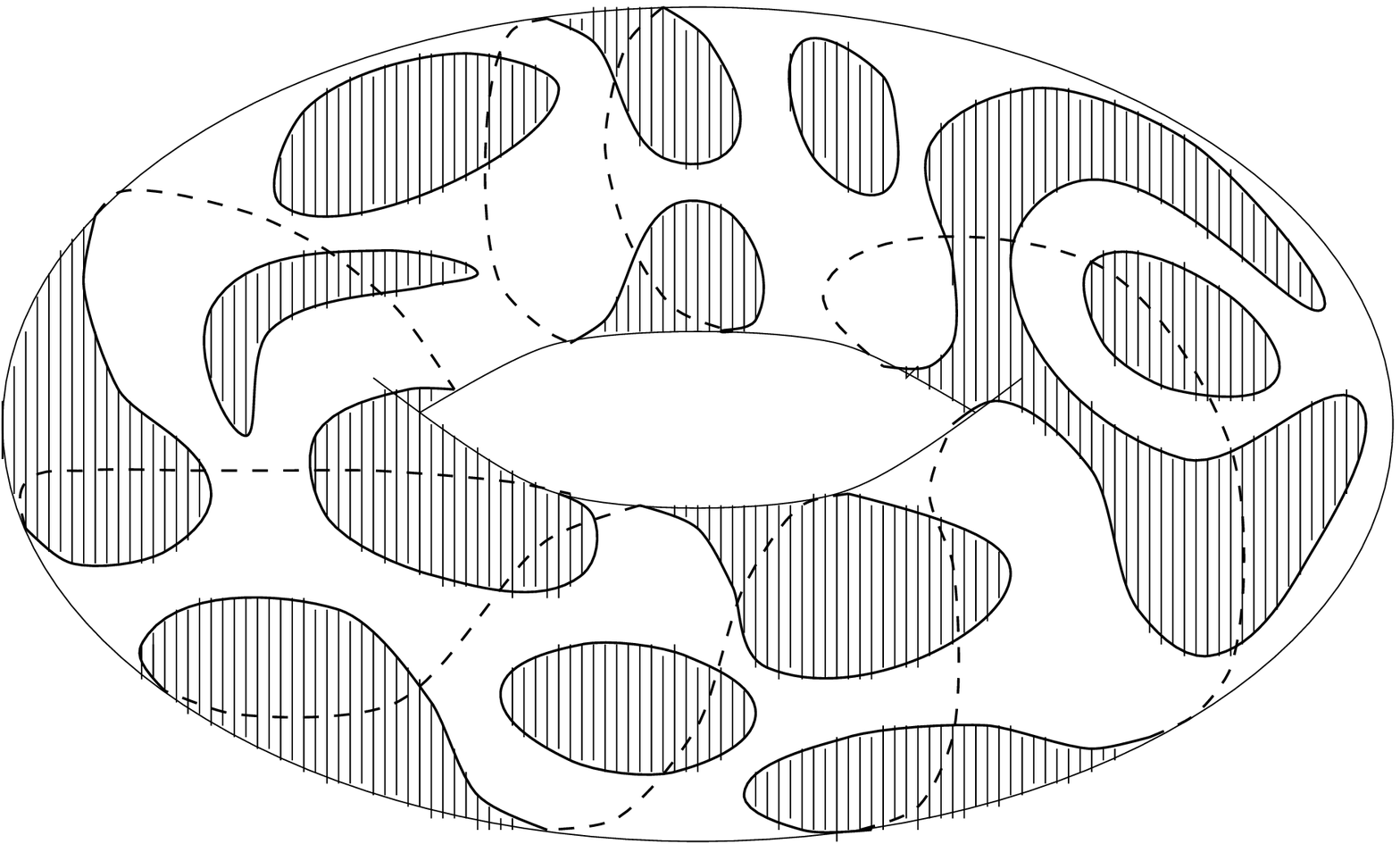}}
\noindent{\ninepoint\sl \baselineskip=8pt {\bf Figure 2}: {\sl
The corresponding Riemann surface. The holes in the ribbon graph
in Figure 1
are filled with shaded regions, all of which have the topology
of the disk.}}

\bigskip
\bigskip

In string theory, various $U(N)$ gauge theories can be realized on D branes.
There, the ribbon graphs ``come to life'' as open string worldsheets
with holes ending on D branes.  In this setup, the string coupling
$\lambda_s$ is identified
with the gauge coupling as
$$\lambda_s=g_{{\rm YM}}^2.$$
To interpret `t Hooft duality in the context of string 
theory, $F_g(t)$ in
\thooft\ should
be viewed as a closed string genus $g$ amplitude which depends
on some modulus in the target space given by $t=g_{{\rm YM}}^2 N=
\lambda_s N$.  

Note that from the expression \fg\ for the 't Hooft expansion,
one would conclude that
$F_g(t)$ is a smooth analytic function of $t$ at $t=0$.  
However this raises a puzzle:  Typically in large
$N$ dualities, $t$ is interpreted as a size of a geometric object 
in the target space and
$t\rightarrow 0$ leads to vanishing cycles, which in turn signal the breakdown
of closed string perturbation theory and divergences of some amplitudes.  
On the other hand, from the dual gauge description, this lack
of smoothness seems to be in conflict with the expectation that gauge theory
perturbation should be reliable in this limit.  This puzzle arises
for any large $N$ dualities, including AdS/CFT correspondences.  We will
see how this puzzle is resolved in the case of the large $N$ duality
for the Chern-Simons theory.

Since we believe our general strategy for a worldsheet derivation 
of large $N$ duality has a wider
range of applicability, here we give its outline
without specializing to the case of the Chern-Simons gauge theory.  
The basic outline of the idea
was suggested in \gopva, and we will
make it precise in the present paper.

  That a worldsheet derivation could exist
in string perturbation theory is natural since the closed string
coupling constant $\lambda_s$ is
kept small.  On the other hand, $t=\lambda_sN$ is not  necessarily
taken to be small.  This means that 
infinitely many holes can contribute to the amplitudes.  This is
very much like how we deform string backgrounds in closed string 
perturbation theory.
There, we add marginal 
perturbations $t\int d^2z V(z,{\overline z})$ to the
worldsheet action. We can compute worldsheet amplitudes by
expanding it in powers of $t$, which counts the number of
insertions of $V$ and we may find it necessary
to take into account 
infinitely many insertions of $V$. Of course
this is still compatible with string perturbation theory,
which is an expansion in $\lambda_s$.
That D-branes may
be incorporated in such a manner in a closed string theory
was also proposed in early days of D-branes \ref\po{
M.~B.~Green and J.~Polchinski,
``Summing over world sheet boundaries,''
Phys.\ Lett.\ B {\bf 335}, 377 (1994);
{\tt arXiv:hep-th/9406012}.
}.

The basic idea we shall use in a worldsheet derivation
of the large $N$ duality is the following:  
We start with a ``good'' description of the worldsheet theory
at $t=0$ from the perspective of the proposed closed string dual.  
By a  ``good'' description we mean 
the one in which the worldsheet quantum field 
theory is well-defined and can be used to describe physics
near $t=0$.
In particular we assume that the
expected breakdown of string perturbation theory and divergence
of string amplitudes
can be traced back to the emergence of an extra non-compact branch 
(or more specifically, the
emergence of a non-compact region in the field space) in the
otherwise perfectly well-defined quantum field theory in two
dimensions.  Such examples
have already been encountered in singular
limits of string compactification \lref\wittencom{E.
Witten, ``Some comments on string dynamics,'' {\tt arXiv:hep-th/9507121}.}
\lref\oogva{H. Ooguri and C. Vafa, ``Two-Dimensional 
Black Hole and Singularities
of CY Manifolds,'' Nucl. Phys. B {\bf 463}, 55 (1996); {\tt arXiv:hep-th/9511164}.}
\refs{\siw,\wittencom,\oogva }.
Let us call the bulk of the field space by $H$ and the new emerging branch 
at $t=0$ by $C$. (The choice of the terminology is motivated by the fact
that, in the Chern-Simons case, $H$ represents the Higgs branch and
$C$ the Coulomb branch of the worldsheet theory.)   
Thus, when we integrate over field configurations on a fixed 
Riemann surface, there may be regions of the surface which 
are in the $H$ phase and there may be others in the $C$ phase. 
The $t$ dependence
of the amplitudes will be captured entirely by the $C$ branch.

In principle, any configuration of $H$ and $C$ domains
can contribute to the closed string amplitudes. However,
in order to prove the correspondence between
gauge theory and string theory,
 we need to establish that Riemann surfaces
with mixed phases do not contribute to the topological
string amplitudes, unless  
each connected component of the $C$ domain has the topology 
of the disk. This is to avoid configurations such as those
shown in Figure 3, which do not correspond to ribbon
graphs in the 't Hooft expansion:

\bigskip
\centerline{\epsfxsize 2.5truein \epsfysize 3.5truein\epsfbox{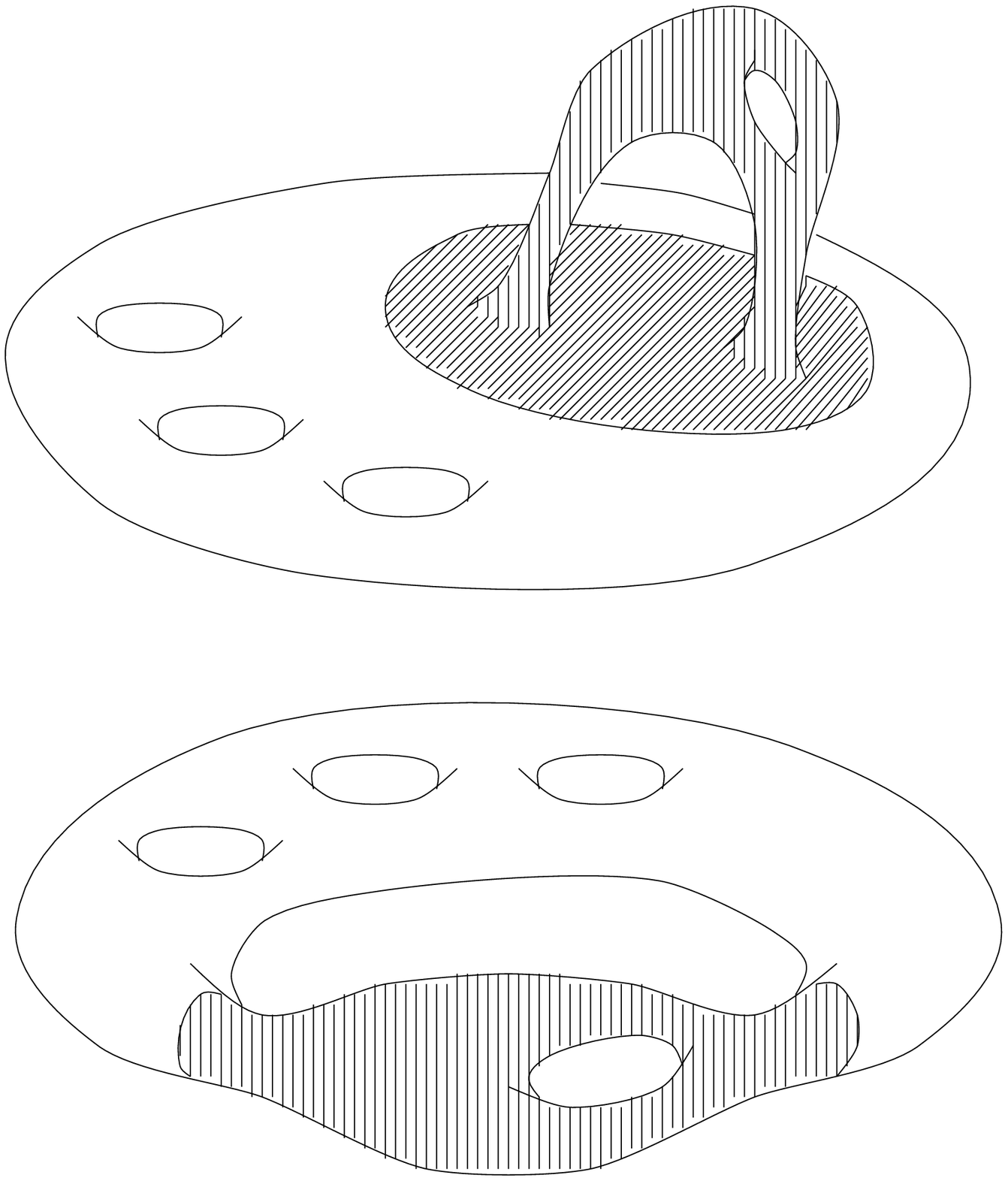}}

\medskip

\centerline{\epsfxsize 3truein \epsfysize 1.8truein\epsfbox{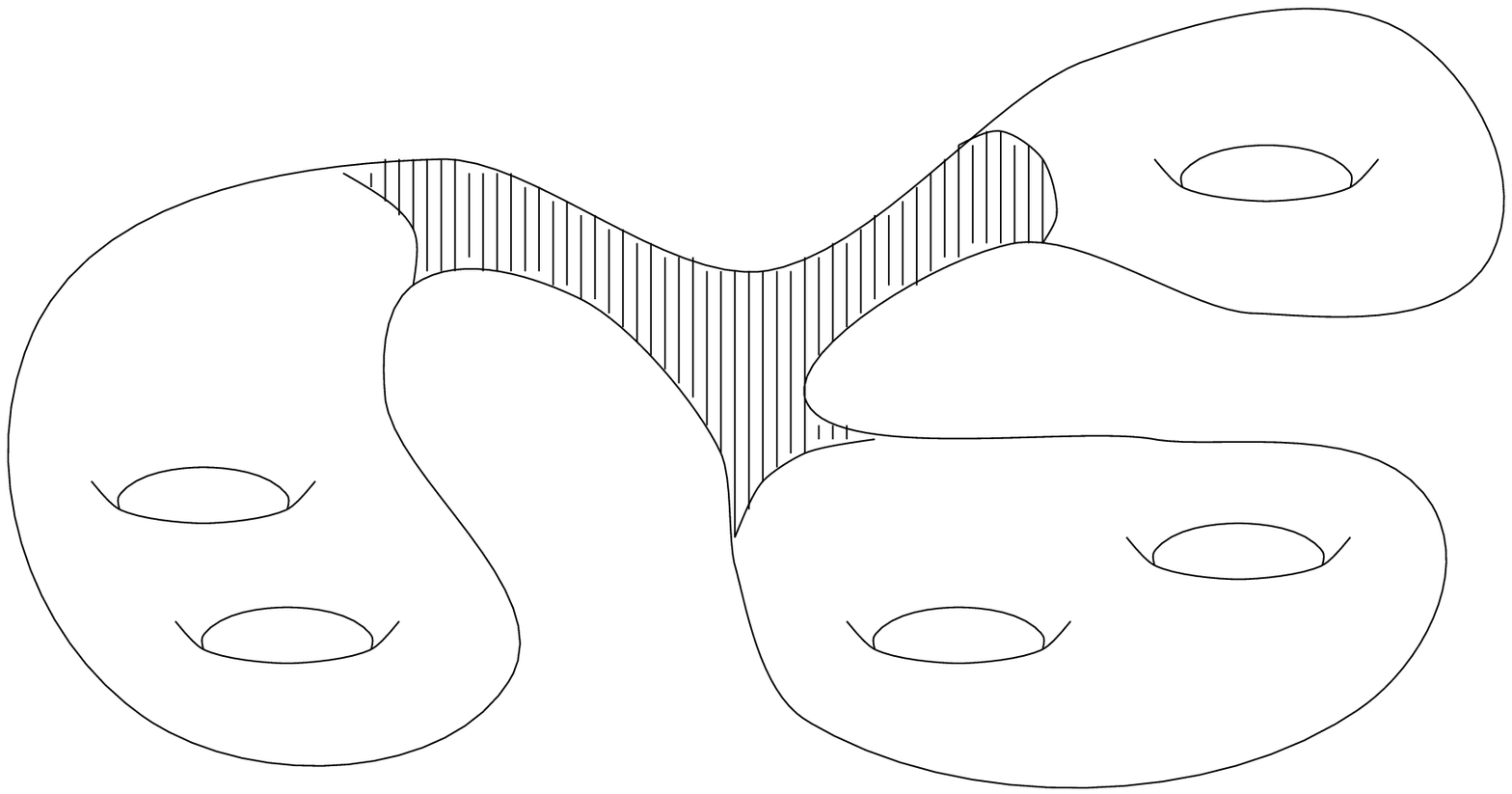}}

\noindent{\ninepoint\sl \baselineskip=8pt {\bf Figure 3}: {\sl
We need to show that worldsheets with both $C$ and $H$ phases, 
such as those shown here, do not contribute to the topological
string amplitudes, 
except when every $C$ domain has the topology
of the disk as shown in Figure 4. 
The $C$ domains are represented in the shaded
regions in the figures.}}
\bigskip

\noindent
Instead we show that the only mixed configurations which 
contribute to the topological string amplitudes are of
the form in Figure 4:

\bigskip
\centerline{\epsfxsize 3truein \epsfysize 1.5truein\epsfbox{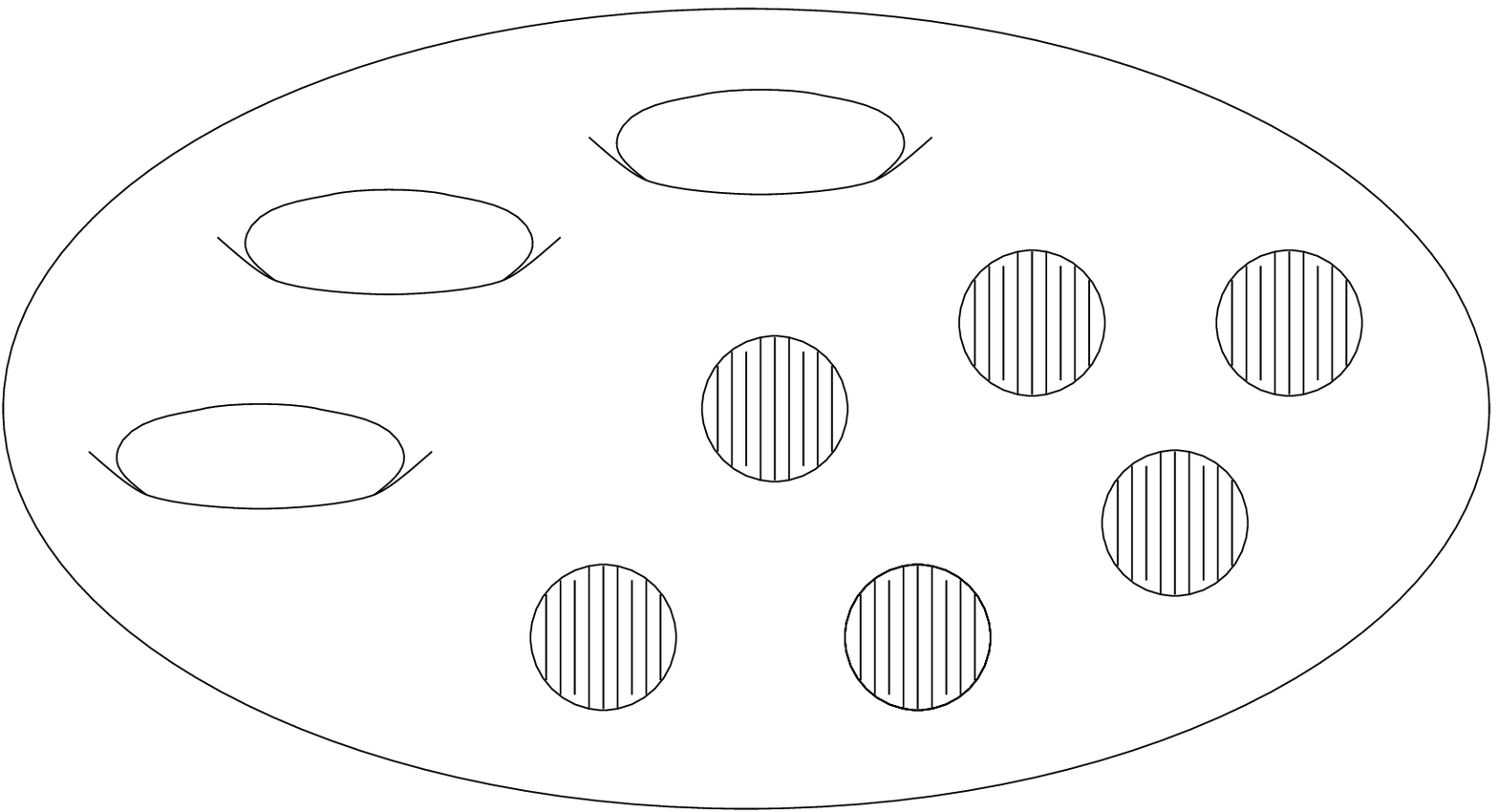}}
\noindent{\ninepoint\sl \baselineskip=8pt {\bf Figure 4}: {\sl
A worldsheet with an arbitrary number of $C$ domains with the
topology of the disk can contribute. 
We will show that each $C$ domain contributes with
the factor of $t=g_{{\rm YM}}^2 N$.}}
\bigskip

To reproduce the 't Hooft expansion \fg, each $C$ domain of 
the disk topology should contribute
by a factor of $t=g_{{\rm YM}}^2 N$.   Furthermore we need
that the fields
dynamical in the $H$ phase become non-dynamical in the $C$ phase
($i.e.$ become infinitely massive in the IR in the $C$ phase), so
that we obtain suitable D-brane like conditions
imposed on the $H$ fields on the boundary.  
This precisely reproduces
the structure of 
the 't Hooft expansion \thooft\ of the gauge theory
if we regard $C$ domains as holes of the ribbon graph,
except that each of them gives rise to the necessary 
factor of $t$. To summarize, in this scenario, 
disks in the $C$ phase can be viewed as holes for some open 
string theory living in the $H$ phase. This will lead 
to a D-brane description and thus
to a dual $U(N)$ gauge theory.

We have, however, two more possibilities:  The full Riemann
surface may be in the $H$ or $C$ branch.

The appearance of a worldsheet in the 
pure $H$ phase is
rather natural. If the whole Riemann surface
is in the $H$ branch, we can consider it as contribution of 
purely closed string loops in the presence of D-branes 
before taking the decoupling limit. The fact that such 
contributions are included in our
computation suggests that the large $N$ duality
can be established in a more general sense.
We believe that that our derivation should apply not only
to the gauge theory/closed string duality in the strict
low energy limit of string theory but also to a more 
general setup of the equivalence between a D-brane
configuration and a closed string background, where the gravity is not
decoupled in the D-brane side and the ``near-horizon
limit'' is not taken in the closed string side. 
Of course if we take a decoupling
limit, where the gauge theory decouples from gravity,
the contribution from the entire Riemann surface being
in the $H$ branch should vanish.

 On the other hand, the
case when the whole Riemann surface is in the $C$ branch, depicted
in Figure 5 below, might appear
as surprising from the gauge theory side, as it does not correspond
to any perturbative `t Hooft diagram.  This must thus correspond
to some non-perturbative terms on the gauge theory side.  The fact that
the existence of $C$ branch corresponds to the breakdown of closed
string amplitudes suggests that the contributions
coming from Riemann surfaces which are entirely in the $C$ branch
 are not smooth in the limit 
$t\rightarrow 0$.  From the perspective
of gauge theory, this appears as the case where ``the entire
Riemann surface is a hole''!  Thus our proof points
out that the 't Hooft expansion of the gauge theory may
miss important terms in the closed string side, which correspond
to some non-perturbative terms in the gauge theory.

\bigskip
\centerline{\epsfxsize 2.8truein \epsfysize 1.7truein\epsfbox{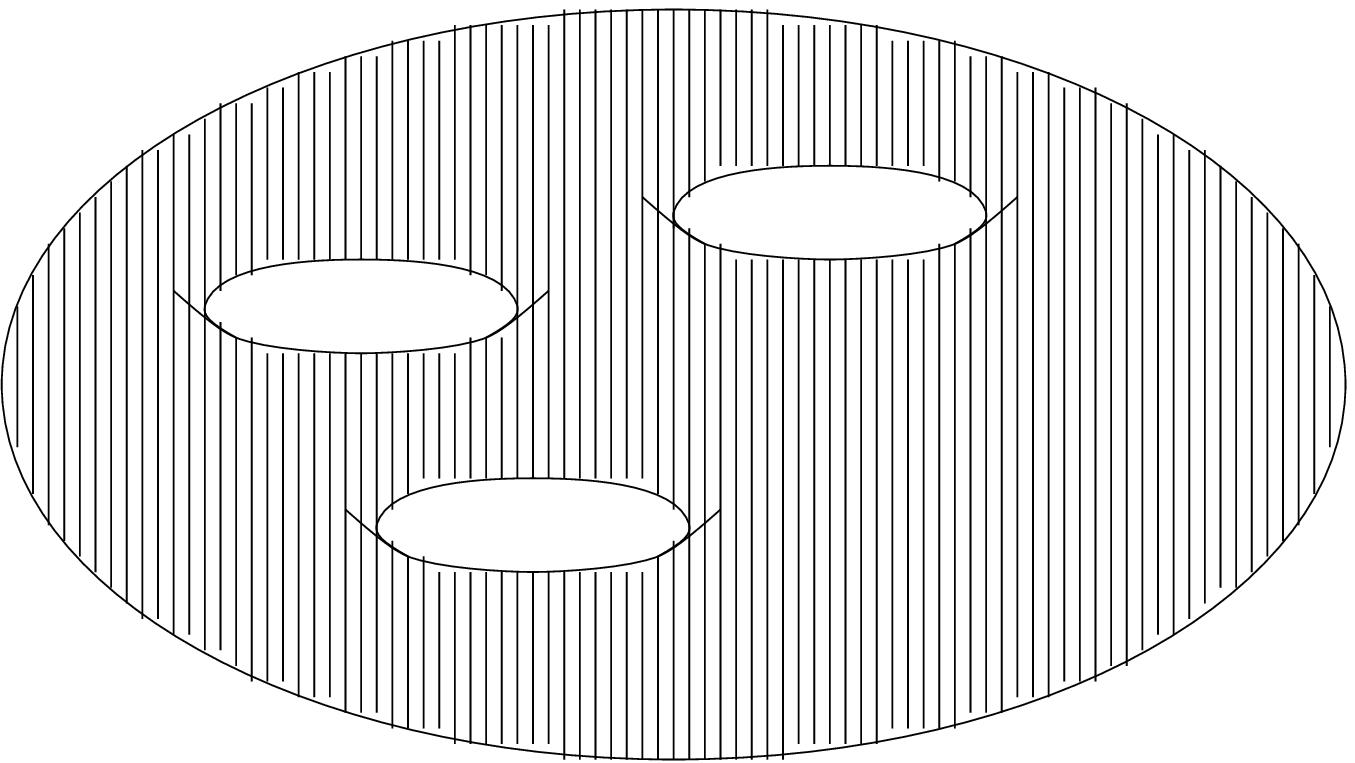}}
\noindent{\ninepoint\sl \baselineskip=8pt {\bf Figure 5}: {\sl
The entire worldsheet can be in the $C$ domain. This does not 
correspond to any perturbative 't Hooft diagram and therefore
must represent some nonperturbative effect in the gauge theory. }}
\bigskip

In this paper we show how the above general ideas are realized
in a precise quantitative manner for the large $N$ duality
of Chern-Simons theory.
The organization of this paper is as follows:  In section 2 we review
aspects of the Chern-Simons/topological closed string duality.
  In section 3 we review
the linear sigma model description of the worldsheet theory on the
resolved conifold.  In section
4 we present the proof of the duality.  
We also give a brief discussion on the cases
of the $SO(N)$ and $Sp(N)$ gauge groups. 
 We will conclude
this paper in section 5 with summary of our results, 
generalization including Wilson loop observables, and 
discussion on implications of our results for other 
large $N$ dualities.

\bigskip
\bigskip

\newsec{Large $N$ Chern-Simons Duality}

The $U(N)$ Chern-Simons gauge theory at level $k$ on $S^3$ was
conjectured in \gopva\ 
to be equivalent to topological closed strings on resolved conifold, with
the dictionary that the string coupling constant
is given by $\lambda_s=i/(k+N)$ 
and the complexified K\"ahler class $t$ of the
resolved conifold is $t= iN/(k+N)$. Since $t$ is pure imaginary,
the $S^2$ of the conifold
geometry has zero size but there is a non-vanishing 
NS-NS two-form field, which makes the string theory non-singular.
We will often refer to $t$ as the size of the $S^2$. 
The motivation for the conjecture comes from the fact that both
sides can be viewed as topological strings near the conifold. 
The $U(N)$ Chern-Simons
theory can be viewed as target string field theory
for topological A-model on the deformed conifold
where there are $N$ D-branes wrapping around 
$S^3$ \wittenjones. The gravity dual
is the topological string on the resolved conifold.  Thus, very much
in the spirit of large $N$ dualities, there is a topological transition
where the $S^3$ cycle containing D-branes shrinks and there emerges
a dual cycle $S^2$.  In fact, since the Chern-Simons theory is topological,
the partition function does not depend on the volume of the $S^3$, so we can
take the $S^3$ size to zero.  On the other hand, since
$t$ is pure imaginary, the resolved conifold geometry in the closed
string side is also at zero size.  So, geometrically speaking,
the duality is the conversion of branes to NS-NS $B$-field, as shown
in Figure 6 below, without touching
the underlying geometry.

\bigskip

\centerline{\epsfxsize 4.3truein \epsfysize 4.2truein\epsfbox{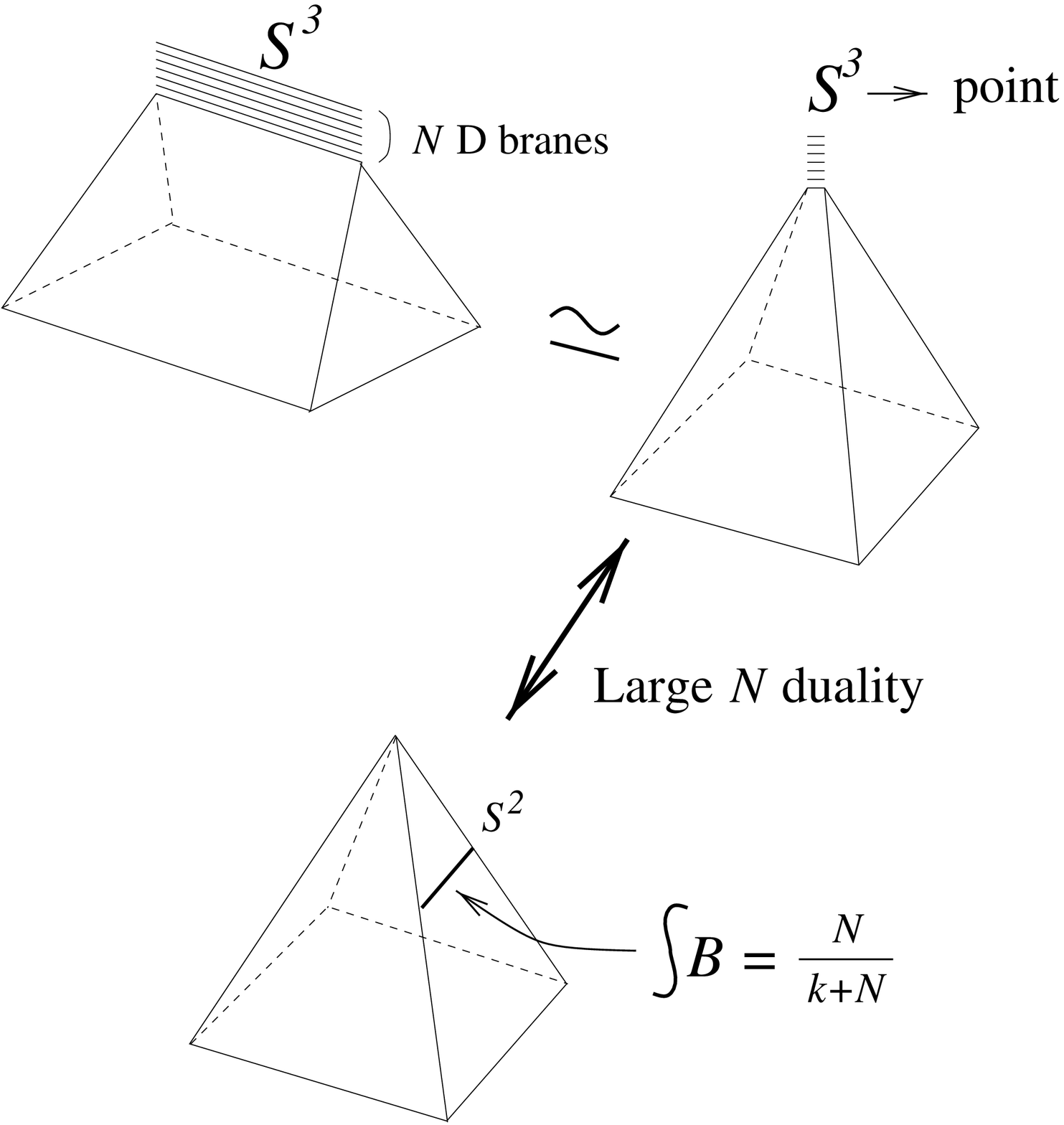}}
\noindent{\ninepoint\sl \baselineskip=8pt {\bf Figure 6}: {\sl
The Chern-Simons theory can be realized as the topological open string
theory on D branes wrapping on the $S^3$ of
the deformed conifold. Topological string amplitudes are independent
of the size of the $S^3$, so we can shrink it to zero. Thus the
large $N$ duality is the statement that $N$ D branes on the
conifold can be replaced by the $B$-field of the amount $g_{{\rm YM}}^2 N$. 
}}

\bigskip

Many checks have been
performed for this conjectured duality including highly non-trivial
tests involving Wilson loop operators to all order in the
gauge coupling constant
\lref\oov{H.~Ooguri and C.~Vafa,
``Knot invariants and topological strings,''
Nucl.\ Phys.\ B {\bf 577}, 419 (2000);
{\tt arXiv:hep-th/9912123}.
}
\lref\mala{M.~F.~Labastida and M.~Marino,
``Polynomial invariants for torus knots and topological strings,''
Commun.\ Math.\ Phys.\  {\bf 217}, 423 (2001);
{\tt arXiv:hep-th/0004196}.
}
\lref\lin{P.~Ramadevi and T.~Sarkar,
``On link invariants and topological string amplitudes,''
Nucl.\ Phys.\ B {\bf 600}, 487 (2001);
{\tt arXiv:hep-th/0009188}.
}
\lref\lmv{J.~M.~Labastida, M.~Marino and C.~Vafa,
``Knots, links and branes at large $N$,''
JHEP {\bf 0011}, 007 (2000); {\tt arXiv:hep-th/0010102}.
} 
\refs{\oov, \mala, \lin, \lmv}.\foot{The original conjecture was 
for the gauge group $SU(N)$, 
but it was pointed out in \ref\marv{M.~Marino and C.~Vafa,
``Framed knots at large $N$,''
{\tt arXiv:hep-th/0108064}.
} that the duality can be stated in a more natural form
in the context of $U(N)$ gauge theory.} 
It is clearly desirable to go beyond gathering evidence for 
large $N$ dualities and actually prove them. 
The Chern-Simons theory/topological closed string duality 
was embedded in the context of the type IIA string theory
in  \ref\vaug{C.~Vafa,
``Superstrings and topological strings at large $N$,''
J.\ Math.\ Phys.\  {\bf 42}, 2798 (2001);
{\tt arXiv:hep-th/0008142}.
}. It is known \oov\  that 
amplitudes of the Chern-Simons theory can be used to
compute the $F$ terms in the ${\cal N}=1$ gauge theory which
is realized on $N$ D6 branes wrapping
around the $S^3$ in the deformed conifold geometry.
On the other hand, topological closed
string amplitudes give corresponding terms in the
low energy effective theory of the type IIA string theory 
on the resolved conifold with $N$ units of the Ramond-Ramond 
flux through the $S^2$ \bcov . 
Therefore we have a proof of the Chern-Simons/topological closed string
duality {\it provided} one can show that
the deformed conifold with D6 branes and 
the resolved conifold without brane are related in
the type IIA string theory.
Subsequently it was found that
these two configurations 
are indeed smoothly connected to each other by lifting
 them to $M$ theory on a $G_2$ holonomy manifold
\ref\amv{M.~Atiyah, 
J.~M.~Maldacena and C.~Vafa,
``An $M$ theory flop as a large $N$ duality,''
J.\ Math.\ Phys.\  {\bf 42}, 3209 (2001);
{\tt arXiv:hep-th/0011256}.
}.
The existence and absence of D-branes on these two
configurations are traced to the fact that,
depending on the moduli of the $G_2$ manifold, 
a $Z_N$ action on the $G_2$ manifold does or does 
not have a fixed point on the $S^3$. Moreover one
can show that the $F$ terms in question do not depend on
these moduli. Therefore the $M$ theory duality
implies the Chern-Simons/topological closed string
duality. Various aspects
of the lifting of type IIA configurations
to $G_2$ holonomy manifolds 
have been studied in a number of papers 
\lref\ach{B.~S.~Acharya,
``On realising ${\cal N}=1$ super Yang-Mills in M-theory,''
{\tt arXiv:hep-th/0011089}.
 }
\lref\agv{M.~Aganagic and C.~Vafa,
``Mirror symmetry and a $G_2$ flop,''
{\tt arXiv:hep-th/0105225}.
}
\lref\aw{
M.~Atiyah and E.~Witten,
``$M$ theory dynamics on a manifold of $G_2$ holonomy,''
{\tt arXiv:hep-th/0107177}.
 }
\refs{\ach, \agv, \aw}.

This derivation is fine, except that one is appealing to the
duality between the type IIA theory and $M$ theory compactified 
on a circle, which remains a conjecture. The aim of this paper 
is to give a self-contained proof of the gauge theory/string theory
 duality 
without making use of the $M$ theory/type IIA duality.
The $g$ loop topological string amplitude is evaluated
by computing the functional integral of the sigma-model
for the resolved conifold and then integrating the resulting
amplitude over the moduli space of genus $g$ Riemann surface.
We will show explicitly that this string amplitude
can be expressed as a sum of Feynman
diagrams in the Chern-Simons gauge theory. 

That there may exist a worldsheet derivation of the Chern-Simons/
topological strings was suggested in \gopva .  The basic
idea advocated there was to start from the closed string side
and go to a regime where the conifold geometry is singular, 
which corresponds
to small 't Hooft coupling in the gauge theory side.  In this limit,
the worldsheet theory acquires Coulomb and Higgs phases, which
can co-exist on a single worldsheet.  The basic proposal
there was that the regions on the worldsheet
which are in the Coulomb phase become the holes from the
perspective of the Higgs phase and that this effectively 
introduces an open string sector in the string theory.  
In this paper, we show that this general idea is correct
and prove that amplitudes in the Chern-Simons theory are 
quantitatively
reproduced in the topological closed string computation.

 For strings
on the resolved conifold, where various aspects
of the worldsheet dynamics have been studied
\lref\ghv{D.~Ghoshal 
and C.~Vafa,
``$c = 1$ string as the topological theory of the conifold,''
Nucl.\ Phys.\ B {\bf 453}, 121 (1995);
{\tt arXiv:hep-th/9506122}.
 } 
\refs{\ghv, \oogva},
it is known that the spectrum becomes
continous without mass gap and that
$g$ loop topological string
amplitudes diverges as $1/t^{2g-2}$ in
the limit $t \rightarrow 0$. This is in line with the expectation
that string theory perturbation breaks down in this limit.
However, as discussed in the introduction, this seems to be
at odds with the structure of the 't Hooft expansion of 
$F_g(t)$ shown in \fg\ and its smoothness as $t\rightarrow 0$. 

It turns out that the exact Chern-Simons amplitudes also
contain terms singular in $t$. That these amplitudes
have such extra terms that are missing in the perturbative
expansion has been noted 
and how they appear from the path-integral
has been studied in \wittenknot\ and \lref\freegom{D. Freed
and R. Gompf, 
``Computer calculation of Witten's three manifold invariant,''
Commun.\ Math.\ Phys.\  {\bf 141}, 79 (1991).
}
\lref\jeff{
L.~C.~Jeffrey,
``Chern-Simons-Witten invariants of Lens spaces and torus bundles, 
and the semiclassical approximation,''
Commun.\ Math.\ Phys.\  {\bf 147}, 563 (1992).
}
\lref\roz{L. Rozansky, ``A large $k$ asymptotics
of Witten's invariant of Seifert manifolds,''
Commun.\ Math.\ Phys.\  {\bf 171}, 279 (1995);
{\tt arXiv:hep-th/9303099}.}
\refs{\freegom,\jeff,\roz}. Similar non-perturbative terms
arise in the context of the emedding of this duality in type IIA
superstrings.  It is important for our derivation
of large $N$ dualities to study these extra terms
which are not captured by the `t Hooft expansion.

\subsec{Nonperturbative terms; Chern-Simons perspective}

Let us first discuss the appearance of these
non-perturbative terms from the perspective of the Chern-Simons
gauge theory \refs{\wittenknot, \freegom , \jeff , \roz }.  The idea is to
consider the prefactors that go into the definition of the path
integral. The $U(N)$ Chern-Simons perturbation theoy on $S^3$ expands
the gauge field near the trivial connection $A=0$.  This connection
has a residual global gauge symmetry corresponding to constant
$U(N)$ gauge transformations.  This implies that the path-integral
must have an extra factor of $1/{\rm vol}_{{\rm CS}}
\left(U(N)\right)$, where the volume ${\rm vol}_{{\rm CS}}$
is to be measured using the normalization defined by the quadratic
term in the Chern-Simons action.  Since this factor is $k/2\pi$, and 
it is quantum corrected to $(k+N)/2\pi$, 
this means that the radius of $U(N)$ is not 1, 
but rather $\sqrt{(k+N)/2\pi}$.
Thus, rescaling the radius to 1, and denoting the corresponding volume
by ${\rm vol}(U(N))$, we see that the
path integral should have a prefactor
\eqn\csful{\exp\left(\cal F\right)
={(2\pi i)^{{1\over 2}{\rm dim}(U(N))} 
\over \left({k+N\over 2\pi}\right)^{{1\over 2}{\rm dim}\left(U(N)\right)} 
{\rm vol} (U(N))}\exp\left({\cal F}_{\rm pert.}\right),}
where ${\cal F}_{{\rm pert}}$ is computable by summing over
the 't Hooft diagrams. The factor $(2\pi i)^{{1\over 2}{\rm dim}(U(N))}$
in the numerator comes from the Gaussian integral formula
$\int dx e^{-{1\over 2 i}x^2} = \sqrt{2\pi i}$.
 
The non-perturbative piece of the amplitude in this case
is then given by
\eqn\nonperc{ {\cal F}_{\rm nonpert}=- \log\left( {\rm vol}
 (U(N))\right) - {1\over 2} {\rm dim}\left( U(N) \right)
\log\left({k+N\over (2\pi)^2 i}\right).}
Let us evaluate this formula. Since $U(N) = U(1) \otimes SU(N)/Z_N$
and since ${\rm vol}\ U(1) = 2\pi \sqrt{N}$ in our normalization
of the kinetic term, we have
\eqn\voluandsu{ {\rm vol}(U(N)) = {2\pi \over \sqrt{N}} 
{\rm vol}(SU(N)).}
The volume of a compact group has been computed in
\ref\macdonald{
I.~G.~ Macdonald, ``The volume of a compact Lie group,''
Invent. Math. {\bf 56}, 93 (1980).}. Applying this
formula to $SU(N)$, one obtains
\eqn\volumeun{
{\rm vol}\left( SU(N) \right) =
 {\rm vol}\left({\rm g}/{\rm g_{{\bf Z}}}\right) \cdot
{\rm vol}(S^3 \times S^5 \times \cdots S^{2N-1}),}
where ${\rm g}$ is the Lie algebra of $U(N)$,
${\rm g}_{{\bf Z}}$ is the Chevalley lattice,
and $S^{2i-1}$ ($i=2,3,\cdots, N$) are unit spheres
with dimensions $(2i-1)$. 
The odd dimensional spheres appear
in the volume formula since 
$SU(N)$ is rational homotopy equivalent to the product
of these spheres. Evaluating this formula, we find\foot{
Since the kinetic term of the Chern-Simons theory uses
the trace in the fundamental representation, the normalization
of the metric on $SU(N)$ differs by a factor $\sqrt{2N}$ 
from that in the mathematical literature. We thank
Y. Hashimoto for useful communication and for having
his unpublished notes available for us.}
\eqn\volume{ {\rm vol}\left(SU(N)\right)
= {\sqrt{N} (2\pi)^{{1\over 2}N^2+{1\over 2}N-1} 
\over (N-1)! (N-2)! \cdots 2! 1!}
={\sqrt{N} (2\pi)^{{1\over 2}N^2 + {1\over 2}N -1} \over G_2(N+1)},}
where $G_2(z)$ is the Barnes function defined by
\eqn\barnes{ G_2(z+1) =\Gamma(z) G_2(z), ~~G_2(1)=1.}
Combining this with \voluandsu , we find
\eqn\volumeun{ {\rm vol}(U(N)) =
{(2\pi)^{{1\over 2}N^2+ {1\over 2}N} \over G_2(N+1)}.}
Substituting this into \csful , we obtain the nonperturbative piece
of the Chern-Simons amplitude as
\eqn\nppiece{
{\cal F}_{{\rm nonpert}} =
{\rm log}\left(e^{{\pi \over 8}i N^2} \left({2\pi \over k+N }
\right)^{{1\over 2}N^2}
{G_2(N+1) \over (2\pi)^{{1\over 2}N} } \right) . }
This precisely reproduces the $k \rightarrow
\infty$ limit of the exact result \wittenknot\ in
the Chern-Simons theory.

Now let us take the large $N$ expansion of this formula. 
Using the Binet integral for the Gamma function 
\eqn\binetgamma{
 \log \Gamma(z) = \left(z-{1\over 2}\right)\log z
- z + {1\over 2} \log 2\pi + 2 \int_0^\infty
{\tan \left( {t\over z} \right) \over
e^{2\pi t} - 1} dt,}
one can derive the asymptotic expansion of the Barnes function as 
\eqn\binet{
 \eqalign{ \log G_2(N+1)  =&
{N^2 \over 2}\log N 
-{1\over 12} \log N -{3\over 4} N^2 + {1\over 2} N\log 2\pi + \zeta'(-1)
\cr
&+\sum_{g=2}^\infty {B_{2g} \over 2g(2g-2) N^{2g-2}}.}}
Recognizing $(k+N)=1/{\lambda_s}$, we obtain 
\eqn\nonperturbative{
\eqalign{{\cal F}_{{\rm nonpert}} = &
{N^2 \over 2} \left( \log \left(2\pi i N\lambda_s \right) 
-{3\over 2}\right)
-{1\over 12} \log \left( N \right) +\zeta'(-1) \cr
&~~~~~
+ \sum_{g=2}^\infty {1 \over N^{2g-2}}
{B_{2g}\over 2g(2g-2)} ,}}
which we can rewrite in terms of $\lambda_s$ and $t=N\lambda_s$ as
\eqn\annon{
\eqalign{ {\cal F}_{{\rm nonpert}} = &
{1\over 2} \lambda^{-2} t^2 \left(  \log \left( 2\pi i
t\right) -{3\over 2}\right)
-{1\over 12} \log \left( 
t \lambda_s^{-1} \right) +\zeta'(-1) \cr
&~~~~~
+ \sum_{g=2}^\infty \lambda_s^{2g-2}
{B_{2g}\over 2g(2g-2)t^{2g-2}}.}}
As we will show in section 4.2, 
this structure of singularity predicted by the prefactors
in Chern-Simons theory exactly agrees with the result
of the topological closed string computation for
worldsheets in the pure $C$ phase.\foot{In order
to check this, one has to recall the ambiguities inherent 
in the definition of closed string partition function 
genus $0$ and $1$.}

\subsec{Nonperturbative terms; type IIA superstring perspective}

These nonperturbative terms also have physical
interpretations in the context of the type IIA 
string theory, which is far more non-trivial.
The Chern-Simons amplitude
compute some $F$-terms in the ${\cal N}=1$ gauge theory realized
on D6 branes wrapping on $S^3$ on the deformed conifold \bcov .  In
particular the planar diagrams of the Chern-Simons theory compute
superpotential terms in the gauge theory involving the gaugino
bilinear superfield $S=Tr \psi^2$.
If $F_{0,h}$ denote the Chern-Simons amplitude
with worldsheet topology of genus 0 and with $h$ holes, 
one obtains a superpotential
\eqn\pertexp{
W_{{\rm pert}}(S)=N\sum_{h=1}^\infty F_{0,h}S^{h-1}=N\sum_{n\not =0}
\ (S+in)\
{\rm log}(S+in).}
On the other hand, 
the exact answer for the Chern-Simons gauge theory is 
known, and the leading term in the large $N$ expansion yields
\eqn\exact{ W_{{\rm exact}}(S)=N\sum_{n} \
(S+in)\ {\rm log}(S+in)=N {\partial F_0\over
\partial S},}
where $F_0$ is the genus $0$ closed string amplitude on the resolved
conifold with Kahler class $t=S$. Compared with \pertexp, 
the sum here includes the $n=0$ term,
which corresponds to $N S \log S$ term. This term is absent
from the perturbative expression \pertexp\ but is a part
of the exact Chern-Simons amplitude, as discussed above.  

The splitting of the exact Chern-Simons partition function \exact\
to perturbative and non-perturbative pieces has a
physical interpretation in type IIA superstring.  The
low energy effective theory of $N$ D6 branes wrapping $S^3$ 
is the ${\cal N}=1$ Yang-Mill theory.  The superpotential
$W(S)$ receives stringy corrections, and the perturbative
string computation is caputured in $W_{{\rm pert}}(S)$.  
However, there are also non-perturbative contributions:  
Gauge theory instantons generate a superpotential for $S$.  
In fact to leading order in IR, which
means the small $S$ regime, the measure induced
potential $NS \log S$ \ref\venyan{G. Veneziano
and S. Yankielowicz, 
``An effective Lagrangian for the pure ${\cal N}=1$ 
supersymmetric Yang-Mills theory,''
Phys. Lett. B {\bf 113}, 231 (1982).}\
and the $\tau S$ tree term,
with $\tau=g_{{\rm YM}}^{-2} $, dominate the superpotential
and lead to gaugino condensate
$$dW=0\rightarrow d(NS{\rm log}S+\tau S)=0\rightarrow S\sim e^{-\tau/N}.$$
Thus it is not surprising that $S \log S$ is missing from the
pertubation theory:  it should come from axial anomaly
in path integral measure. That they are related to 
fractional instantons is 
reflected in the fact that the condensate is exponential $\tau/N$ which
is $1/N$-th of the action for an instanton.  
The reason that this survives at large $N$ is precisely
because they are fractional, so that they are weighted with $1/N$
and one gets the action as inverse of `t Hooft parameter $1/Ng_{\rm YM}^2
= 1/t$.

Note that the wrapping Euclidean D2 brane, which from the viewpoint
of D6 brane is an instanton, is not an instanton
from the viepoint of the Chern-Simons
gauge theory since it is not localized
to a point on $S^3$.  This is consistent with the fact
that these non-perturbative terms appear in the prefactor
$1/{\rm vol}\left(U(N)\right)$ in the context
of Chern-Simons gauge theory.

Similar contributions exists at each order in $1/N$.  As discussed
in \vaug , each term in the $1/N$ expansion correspond
to some contribution to a superpotential-type term.
Moreover at each order in the $1/N$ expansion, which
corresponds to a fixed genus on the closed string dual, the
superpotential terms can be viewed as contributions from perturbative
string expansions plus one term at each genus which can be presumably
attributed to the contribution of fractional instantons to the
corresponding amplitudes.  If $F_g(S)$ denotes the dual closed string
amplitude at genus $g$ the leading term as $S\rightarrow 0$ is precisely
the term missing from open string/gauge theory perturbations.
For $g \geq 2$, it is given by
$$F_g(S)\rightarrow {B_{2g}\over 2g (2g-2)}S^{2-2g}
~~~~(S \rightarrow 0).$$

We will show that, in the topological closed string side,
these nonperturbative terms correspond to the situation
in which
the entire worldsheet is in the Coulomb phase, $i.e.$ when the
worldsheet as a whole is a ``hole.'' This configuration does not
correspond to any Feynman diagram of the gauge theory,
and it explains why these
terms are not captured in the perturbative gauge theory
computation.  Thus,
both the exact answer in Chern-Simons theory 
and closed string dual 
have these extra terms, and this is perfectly consistent 
with the large $N$ duality.

\newsec{Linear sigma-model description of the worldsheet}

One of the main difficulties in finding a direct proof
of the $AdS$/CFT correspondence has been
a lack of a useful description of string worldsheet
when the 't Hooft coupling in the gauge theory side
is small. In fact, for string theory on the resolved
conifold, there is such a description.
The linear sigma-model \WittenYC\ 
is a good description for the worldsheet even when
the size $t$ of the base $S^2$ is small.

Before describing the linear sigma-model,
it would be useful to start with a brief review of
the conifold geometry. The conifold is a singular
space defined by the equation,
\eqn\coni{   \sum_{i=1}^4 z_i^2 =0 ,}
in ${\bf C}^4$. The K\"ahler form is given by
\eqn\kahler{ \omega = {1 \over 2i} \sum_{i=1}^4 dz_i \wedge d\bar{z}_i,}
restricted on \coni .
The space has a singularity at $z_i=0$, and
it can be made smooth either by
deformation of complex
structure or by small resolution. The deformation of the conifold
is described by the equation,
\eqn\deformedconi{
  \sum_{i=1}^4 z_i^2  = \mu .}
Without loss of generality, we can assume that $\mu$ is
real and positive. 
The resulting space is nothing but the cotangent space of $S^3$. To see
this, we set 
\eqn\phasecoord{z_i = x_i + i p_i.}
The equation can then be expressed
as
\eqn\cotangent{
  \sum_i x_i^2 - \sum_i p_i^2 = \mu,~~~
\sum_i x_i p_i =0.}
The first equation shows that the space contains
an $S^3$ of radius $\mu$ at $p_i=0$, and the second
equation means that $p_i$'s are coordinates of the
cotangent space at $x \in S^3$. Moreover the K\"ahler form
\kahler\ is expressed in these coordinates as
\eqn\phasespace{ \omega =  \sum_{i=1}^4 dp_i \wedge dx_i ,}
which gives the natural symplectic structure on $T^* S^3$.
Thus the space \deformedconi\ with $\mu \neq 0$
is smooth deformation of the singular space
\coni .

The conifold can also be made smooth by small resolution.
It means finding
a smooth space which can be mapped holomorphically
onto the conifold \coni\ except at the singularity
at $z_i =0$. Thus it is not a complex structure
deformation but should be considered as deformation
of the K\"ahler structure.
To describe the process, it is useful to introduce
a new set of complex coordinates $y_i$ ($i=1,\cdots,4$) defined by
\eqn\changecoords{ \eqalign{
  y_1 &= z_1 + i z_2, ~~~y_2 = z_1 - i z_2 , \cr
  y_3 & = z_3 + i z_4, ~~~y_4 = - z_3 + i z_4.}}
The conifold equation \coni\ can then be written as
\eqn\conitwo{ y_1 y_2 - y_3 y_4 = 0.}
Introducing the two sets of coordinates $(y_1, y_2, y_3, z)$
and $(w, y_2, y_3, y_4)$ defined by
\eqn\blowup{ \eqalign{
(y_1,y_2,y_3,y_4) &= (y_1, y_2, y_3, zy_1) \cr
  & = (wy_4, y_2, y_3, y_4),}}
the equation \conitwo\ becomes
\eqn\conithree{ y_1 (y_2 - z y_3) = 0}
in the $(y_1, y_2, y_3, z)$ coordinates and
\eqn\conifour{ y_4(w y_2 - y_3) = 0}
in the $(y_2, y_3, y_4, w)$ coordinates.
This suggests considering
two smooth spaces
\eqn\smoothunion{
U_1 = \{ \ (y_1, y_2, y_3, z) \ | ~~ zy_2-y_3=0 \ \},
~~~ U_2 =\{ \ (y_2, y_3, y_4, w) \ | ~~ y_2-wy_3=0 \ \} ,}
glued together by $zw=1$ and $y_1=zy_4$. The resulting
space is smooth, and it is mapped onto the conifold \conitwo\
by \blowup . The map is holomorphic except at
$y_i=0$, and this gives the small resolution.

The inverse image of the singularity $y_i=0$
is $z$ and $w$ glued together by $zw=1$, namely
we can regard $z$ and $w$ as coordinates
on the northern and the southern hemispheres
of $S^2$. In this way, the conifold singularity is
resolved by replacing the singularity at $y_i=0$ by the $S^2$.
The equations $y_4=zy_1$ and
$y_2=z y_3$ define a sum of line bundles
${\cal O}(-1) \oplus {\cal O}(-1)$ over the $S^2$,
where $(y_1, y_3)$ are regarded as coordinates
of the fiber over the northern hemisphere and
$(y_2, y_4)$ are over the southern hemisphere
of the $S^2$.

To construct a sigma-model whose target space is the resolved
conifold, it is convenient to use the projective
coordinates $(a_1, a_2)$ on the $S^2$  so that $z=a_1/a_2$.
The north pole is at $a_1=0$ and the south pole is
at $a_2=0$. The coordinates $y_i$'s can then be
expressed as
\eqn\generalsol{ \eqalign{ y_1 = a_1 b_1,~~~y_2 = a_2 b_2 , \cr
                           y_3 = a_1 b_2,~~~y_4 = a_2 b_1.}}
In these coordinates, the $S^2$ is located at
$b_1=b_2=0$.  The variables $a_i, b_i$ are arbitrary, but
there is a gauge symmetry
\eqn\gauge{ (a_i, b_i) \rightarrow (\zeta a_i, \zeta^{-1} b_i),
~~~\zeta \in {\bf C} \backslash \{ 0 \},}
which keeps $y_i$'s invariant. Thus the resolved conifold
can be constructed as the K\"ahler quotient of
$(a_1, a_2, b_1, b_2)$ by the action of \gauge .

This geometric construction is translated into the field
theory language as follows. We consider a linear sigma-model
in two dimensions with ${\cal N}=2$ supersymmetry
consisting of four chiral multiplets,
$A_i, B_i$ $(i=1,2)$, whose lowest components are $a_i, b_i$
in the above paragraph,
and one vector multiplet $V$ for the $U(1)$ gauge symmetry
to enforce the gauge symmetry \gauge .
The four charged chiral fields carry
 $U(1)$ charges $+1$ and $-1$ for $A$ and $B$ respectively.

Let us look at the potential for the scalar fields.
The
kinetic terms $\bar{A}_i e^{2V} A_i + \bar{B}_i e^{-2V} B_i$
gives
\eqn\kin{ V_{kin} =2|\sigma|^2 \left( |a_1|^2 + |a_2|^2 + |b_1|^2
+ |b_2|^2 \right)
      - D \left( |a_1|^2 + |a_2|^2 - |b_1|^2 - |b_2|^2 \right),}
where $e$ is the $U(1)$ gauge coupling constant on the worldsheet, 
$\sigma$ is the complex scalar field, and
$D$ is the auxiliarly field in the vector multiplet $V$.
Note that both $e$ and $\sigma$ have dimensions of mass
on the worldsheet.
The gauge kinetic energy contributes ${-1 \over 2 e^2} D^2$.
In addition, we can add the Fayet-Illiopoulos term
$-r D$ and the theta term $i{\theta \over 2\pi} F$ for
the gauge field strength $F$. Eliminating the auxiliarly field
$D$ by its equation of motion, we obtain
\eqn\potential{ V=
2|\sigma|^2 \left( |a_1|^2 + |a_2|^2 + |b_1|^2 + |b_2|^2 \right) +
{e^2\over 2}
\left( |a_1|^2 + |a_2|^2  - |b_1|^2
- |b_2|^2
- r \right)^2.}

If $r\neq 0$, solutions to $V=0$ are given by
$\sigma = 0$ and $a_i, b_i$  obeying
the constraint
\eqn\abconstraint{|a_1|^2 + |a_2|^2
- |b_1|^2 - |b_2|^2  = r ,}
modulo the gauge symmetry
\eqn\residualgauge{ a_i \rightarrow e^{i\theta} a_i,
b_i \rightarrow e^{-i\theta} b_i.}
The resulting space
is the resolved conifold. To see this, suppose $r > 0$. (For $r < 0$,
one can just exchange the roles of $a_i$ and $b_i$ below.)
The equation \abconstraint\ says that the space contains a
minimum $S^2$ of radius $\sqrt{r}$ at $b_i=0$. Because of
\generalsol , this $S^2$ is located at $y_i=0$, the
location of the singularity of the conifold. The gauge symmetry
\gauge\ is partially fixed by \abconstraint, with
the residual gauge symmetry manifest in \residualgauge .

When $r=0$, the space of solutions to $V=0$ acquires
an additional branch where $\sigma \neq 0$ and
$a_i, b_i=0$ \WittenYC . If $\theta \neq 0$, the theta term
induces a constant electric field \ColemanUZ ,
\eqn\constem{ F = {e^2 \over 2\pi} \tilde{\theta},}
where $\tilde{\theta} \equiv \theta \ ({\rm mod} \ 2\pi {\bf Z})$
and $|\tilde{\theta}| < \pi$. This gives rise to
the vacuum energy density
\eqn\vacen{ E_{vacuum} = {e^2 \over 2} \left({\tilde \theta
\over 2\pi}\right)^2}
for this branch. In addition, if $r \neq 0$, it gives an
additional contribution of ${e^2 \over 2}r^2$ when
$\sigma$ is large. (This is obtained by setting $a_i, b_i=0$
in \potential .) The vaccum energy density for large $\sigma$
can therefore be expressed as
\eqn\vac{ E_{vacuum} = {e^2 \over 2}\left[ r^2 + \left({\tilde \theta
\over 2\pi}\right)^2\right].}
Thus, in the limit $r \rightarrow 0$ and $\theta \rightarrow 0~
({\rm mod} \ 2\pi {\bf Z})$, this branch is degenerate with the
branch with $\sigma=0$.

Classically, in this limit we would expect two branches: the
Coulomb branch $C$ where $|\sigma | \not=0$ which implies that
$a_i,b_i$ are massive and the Higgs branch where $|\sigma|=0$ but
the $a_i,b_i$ are not zero. In the Higgs branch, $\sigma$ field 
is massive and the $U(1)$ gauge symmetry is spontaneously broken.
It is known that, in the quantum theory these two branches
are smoothly connected.  In the infrared limit,
one expects that, even though the two branches are still connected,
the region that interpolates between them stretches out infinitely long.
 For finite $e^2$, wave functions 
can spread from the Higgs branch to the Coulomb branch.
However the distance between these branch 
becomes infinite in the IR limit $e^2 \rightarrow \infty$
and thus the two branches decouple. 
It is believed that the Higgs branch theory is obtained
by the straightforward $e^2 \rightarrow \infty$ limit 
of the linear sigma model, whereas
the decoupled Coulomb branch theory is obtained by keeping
the combination $\widetilde \sigma = |e|^{-1} \sigma$ finite in
taking the limit and by performing the functional integral over
$\widetilde \sigma$ rather than over $\sigma$
\lref\WittenYU{
E.~Witten,
``On the conformal field theory of the Higgs branch,''
JHEP {\bf 9707}, 003 (1997);
{\tt arXiv:hep-th/9707093}.
}
\lref\AharonyDW{
O.~Aharony and M.~Berkooz,
``IR dynamics of $d = 2$, ${\cal N} = (4,4)$ 
gauge theories and DLCQ of 'little  string theories',''
JHEP {\bf 9910}, 030 (1999); {\tt arXiv:hep-th/9909101}.
}  
\refs{\WittenYU , \AharonyDW}.
This Coulomb
branch being disconnected from the Higgs branch plays no role 
in this paper.
In the following, we will focus on the Higgs branch
theory since its infrared limit is the non-linear sigma-model
on the resolved conifold. However, in the Higgs branch,
there is a region where one connects to the Coulomb branch.  This
is the infinitely long throat which used to be a transition region to the
Coulomb branch.  This throat region, where $|\sigma|\gg 1$,
can be viewed as a ``Coulomb domain''
of the Higgs branch. We will refer to it as the $C$ domain
or $C$ branch.  This should not cause any confusion as the IR disconnected
Coulomb branch plays no role in the string theory studied in this
paper.

We will find it useful to note that the dependence of the linear
sigma-model action on $r$ and $\theta$ can be
expressed as a superpotential
\eqn\twistedpotential{  W=  t \Sigma,}
where
\eqn\whatt{ t= {\theta \over 2\pi} + i r,}
and $\Sigma$ is a twisted chiral field defined
as the field strength of the vector multiplet,
\eqn\fieldstrength{ \Sigma = \bar{D}_+ D_- V.}

In this paper, we are interested in the $A$-type topological
twist of the sigma-model. Field configurations invariant
under the topological BRST symmetry are given by
holomorphic maps from the worldsheet to the resolved
conifold, $i.e.$,
worldsheet instantons. One can show \WittenYC\ that the instanton
actions for such configurations
 are given by
\eqn\instanton{ S_{instanton} = -2\pi i t N,}
where $N$ is the instanton number. Thus we can identity
$t$ as the complexified K\"ahler modulus of the resolved
conifold.
When $t$ is non-zero (with ${\sl Im}\ t \geq 0$),
 topological string amplitudes can be expresed as a sum
over worldsheet instantons weighted with the factor
$e^{2\pi i t N}$. As we take the conifold limit $t \rightarrow 0$,
the sum over the instantons may diverge. In the linear
sigma-model description, the divergence occurs since
the non-compact branch with $\sigma \neq 0$ opens up
in the limit.

\newsec{Derivation of the large $N$ duality}

In this section, we will present a detailed derivation of
the equivalence between the Chern-Simons gauge theory
on $S^3$ and the topological closed string on the resolved
conifold. 

To compute a topological string amplitude, 
one starts with the topological A model, $i.e.$ the A-type topological
twist of the linear sigma-model described in section 2, 
evaluate a certain correlation function of the model,
and integrate it over the moduli space of Riemann surfaces.
For example, the $g$ loop topological string partition function is
given by\foot{We are using the symbol ${\cal F}_g$ to denote the
string amplitude, to distinguish it from the corresponding
gauge theory amplitude $F_g$.}
\eqn\gloop{
  {\cal F}_g = \int_{{\cal M}_g} 
\langle \prod_{i=1}^{3g-3} (\eta_i, G_L^-) 
(\bar \eta_i, G_R^-) \rangle,}
where ${\cal M}_g$ is the moduli space of genus-$g$
Riemann surfaces, $G_L^-$ and $G_R^-$ are 
${\cal N}=2$ supercurrents which have conformal dimensions
(2,0) and (0,2) after the topological twist, 
$\eta_i$ and $\bar \eta_i$ are Beltrami 
differentials on the Riemann surface ${\cal S}$, and 
$$(\eta_i, G_L^-) = \int_{{\cal S}} dz^2 \eta_i(z) G_L^-(z).$$

When the K\"ahler modulus $t$ of the resolved conifold is
large, we can express the topological string amplitude
as a sum of worldsheet instantons, $i.e.$, a sum of holomorphic 
maps from genus $g$ surface to the resolved conifold. 
We will show that, when $t$ is small, the same amplitude 
can be expressed as an asymptotic expansion in powers of $t$, and 
each term in the expansion can be idenfitied as a sum 
over Feynman digrams of the Chern-Simons gauge theory 
drawn as genus $g$ ribbon graphs with a fixed number of holes.

As we saw in \vac , when the complex scalar field $\sigma$ in the
vector multiplet is large, the ground state energy density is
given by 
\eqn\vacagain{ E_{vacuum} = {e^2 \over 2}\ {\rm min}_{n \in {\bf Z}}
\ |t-n|^2.} 
Therefore,
if $t$ is away from integer points, 
the contribution from large value of $\sigma$ is suppressed
in the infrared limit $e^2 \rightarrow \infty$.
On the other hand, as $t\rightarrow 0$, the 
potential barrier for the $C$ branch disappears and
$\sigma$ can grow indefinitely large.
This causes sigma-model amplitudes to diverge in the
limit. To separate the singular part and the regular part
of the $\sigma$ model functional integral,  
we find it useful to divide the Riemann surface ${\cal S}$
into the $C$ domain $U_C$ where $\sigma$ is large and the $H$ domain 
$U_H$ where it is small. 
For definiteness, we introduce a cutoff parameter $\sigma_*$
and define these regions and their boundary $\gamma$ as
\eqn\whatu{\eqalign{ U_C & = \{ z \in {\cal S} 
\ : ~ |\sigma(z)| > \sigma_* \},
\cr
U_H &= 
\{ z \in {\cal S} \ : ~ |\sigma(z)| < \sigma_* \}, \cr
\gamma& = \{ z \in {\cal S} \ :~ |\sigma(z)|=\sigma_* \}. }}
We would like to stress that both the $C$ and
$H$ domains are parts of the Higgs branch theory, as already discussed.
The functional integral can then be performed in the following steps:

\smallskip
\noindent
(1) Integral over $\sigma$ with $|\sigma|>\sigma_*$ and
the rest of the fields in $U_C$. Because of the potential
\potential , the chiral multiplet fields $a_i, b_i$ are
suppressed. This domain describes 
topological string localized at the conifold singularity.

\smallskip
\noindent
(2) Integral over $\sigma$ with $|\sigma|<\sigma_*$ and
the rest of the fields in $U_H$. Because of the cutoff,
the functional integral does not generate singularities
at $t=0$. This domain describes string propagating in 
the smooth part of the conifold.

\smallskip
\noindent
(3) Integral over location of the boundary $\gamma$ of
the two domains.

\medskip

Let us discuss contributions from each integral
separately. We will discuss (3) first. 

\subsec{Integral over $\gamma$}

The location and shape of the boundary
$\gamma$ can be regarded as collective coordinates
of $\sigma$. The task of identifying these
collective coordinates is simplified by the
localization property of topological sigma-model
functional integral. We note that
the gauge 
invariant field strengh $\Sigma$ of the vector multiplet
defined in \fieldstrength\  
is a twisted chiral superfield, and the roles of the 
A and B twists are exchanged when acting on the vector 
multiplet fields. In the topological twisting in question,
both $\partial_z \sigma$ and
$\partial_{\bar z} \sigma$ are BRST trivial.
Therefore the functional
integral over $\sigma$ 
on a Riemann surface of fixed moduli
reduces to a finite dimensional
integral over constant mode of $\sigma$.
This means that the entire
Riemann surface is either in the $C$ domain or in the
$H$ domain. 

The situation is more subtle in the topological
string where we integrate over the moduli space
of Riemann surfaces. As explained in \wittenjones\
and clarified further in \bcov , the scalar field 
$\sigma$ can be non-constant if we take the 
limit in the moduli space where the worldsheet is
infintely elongated along the direction in which $\sigma$ is
varying so that variation of $\sigma$ per unit worldsheet 
length goes to zero. This means that the boundary $\gamma$ 
between the $C$ and $H$ domains can emerge if and only if 
the worldsheet has a long cylindrical region and $\gamma$
cuts through the cylinder, as shown in Figure 7.

\bigskip
\bigskip

\centerline{\epsfxsize 3.7truein \epsfysize 2truein\epsfbox{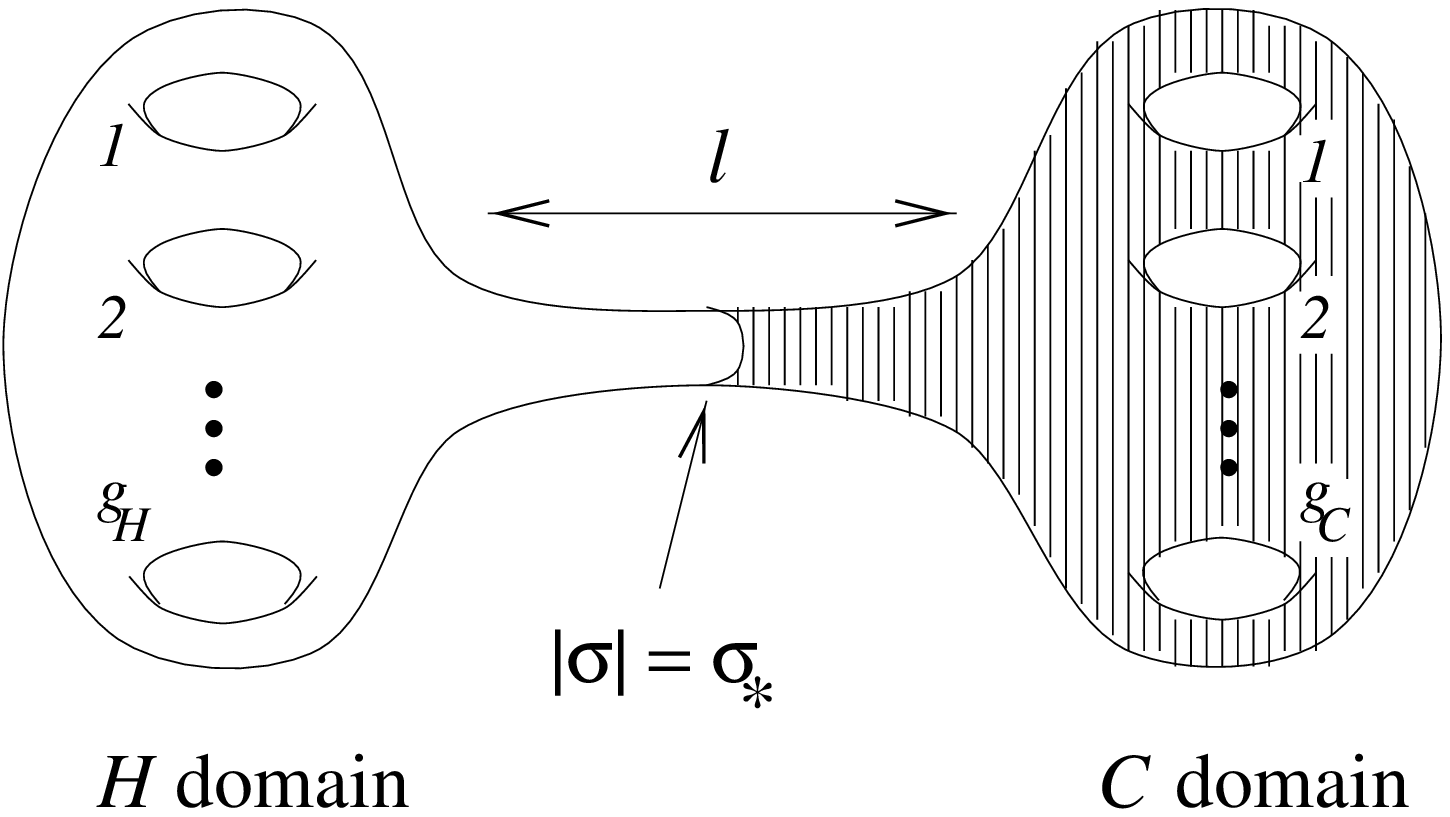}}
\noindent{\ninepoint\sl \baselineskip=8pt {\bf Figure 7}: {\sl
A worldsheet can be separated into the two domains
across a long cylinder.}}

\bigskip
\bigskip

Let us introduce coordinates $(x,y)$ on the cylindrical
region of the worldsheet, where $x$ is a periodic coordinate
around the cylinder, $x \sim x +2\pi$, and $y$ take value
in the interval $0 \leq y \leq l$ for some large $l$.
The cylinder is attached to the rest of the Riemann surface
at the two ends, $y=0$ and $l$. 
Suppose the boundary $\gamma$ is located at $y=y_0$;
the $H$ domain on the left $y< y_0$ and the $C$ domain is
on the right $y > y_0$. By the topological reduction, the $\sigma$ 
field is varying only along the
$y$ direction and stays constant in the $x$ direction. Namely
the vector multilet field reduces to a quantum mechanical model
with $y$ playing the role of imaginary time. 

On the cylinder, we are performing the functional integral
with the constraints that $|\sigma| =\sigma_*$ at $y=y_0$. 
As discussed in the above, the localization of the topological B-model
for the $\sigma$ field implies that, along
the boundary $\gamma $, the functional integral over $\sigma$
 reduces to a finite dimensional integral
over constant value of $\sigma$. We thus have
\eqn\fulld{ \sigma(z) = \sigma_0~~~{\rm at}~~
y=y_0,}
where
\eqn\whatsigmazeo{ \sigma_0= \sigma_* e^{i\varphi}, }
for some fixed $\varphi$.  Of course the
topological sigma model will not fix the phase of $\sigma_0$ and thus
we have to integrate over $\varphi$.
 Since the topological string is B-twisted
in the $C$ domain, the functional
integral in the $C$ domain on the right of $\gamma$ gives an amplitude
which depends holomorphicaly on $\sigma_0$. 
The amplitude from the $H$ domain on the left will turn out 
to be independent of $\sigma_0$. 

In the original functional integral of the linear sigma-model,
the value of $\sigma$ at $y=y_0$ is arbitrary. The constraint
$|\sigma|=\sigma_*$ then transforms the integral over $|\sigma|$
at $y_0$ into an integral over the location $y_0$ of the
boundary $\gamma$ along the cylinder. Let us evaluate the 
Jacobian for this change
of funcational integral variables $(\sigma, \bar{\sigma})
\rightarrow (y_0, \varphi)$. 
We can write
\eqn\jacobian{
 d\sigma(y_0) \wedge d\bar{\sigma}(y_0) =
 d\sigma_0 \wedge dy_0 \cdot{d\bar{\sigma}\over dy}(y_0).}
The operator $d\bar{\sigma}/dy$ is a marginal operator which
shifts the boundary condition $\sigma=\sigma_0 \rightarrow 
\sigma_0 +\epsilon$.  To see this note that $d{\bar \sigma}/dy$ is the momentum
conjugate to $\sigma$ (viewing $y$ as the
time direction along the cylinder) and so it corresponds to 
$\partial/\partial\sigma$.
 Therefore we can make the change of the functional
integral variables as
\eqn\jacobiantwo{ d\sigma(y_0)\wedge d\bar{\sigma}(y_0)
\sim dy_0 \wedge d\sigma_0  \cdot {\partial \over \partial \sigma_0}.}
Here the derivative operator on the right-hand side
acts on the amplitudes computed in the $C$ and
$H$ branches.
We still have to integrate over the localization
moduli for the B-model, which means the phase of the $\sigma$ field.  Thus
we find that the integral over $\gamma$ effectively turns into 
\eqn\gian{dy_0 \oint d\sigma_0  \cdot {\partial \over \partial \sigma_0},}
where the contour integral is over $|\sigma |=\sigma_*$.  The derivative
will act on the path integral on the Coulomb branch with
the boundary condition $\sigma=\sigma_0$ on $\gamma$.
  Since the topological
B-model amplitude is holomorphic in $\sigma_0$, the precise
location of the above
integral is irrelevant as long as it circles the origin, 
where the topological amplitude may have
singularities. The integral over $dy_0$, together with the
overall length $l$ of the cylinder, is exactly what one expects for the
integration over division of the cyclinder to two regions, as one
would need for the interpretation of $H$ and $C$ regions as independent
string theories, whose amplitudes are evaluated by integrating 
over moduli spaces of Riemann surfaces with boundaries.

\subsec{Contribution from the $C$ domain}

The $C$ domain describes topological string at the conifold
singularity. 
Since the chiral multiplet fields are massive, we can integrate
them out. This gives rise to a linear dilaton coupling for
the vector multiplet, which changes the measure for $\sigma$ \siw. 
In addition, we have the twisted superpotential \twistedpotential .
We will utilize this description to compute amplitudes
in the $U_C$ domain.

A topological string amplitude is given by 
an integral of a topological sigma-model correlation function
over the moduli space of Riemann surface. For example, 
the topological string partition function at $g$ loop
is defined by \gloop .
When the worldsheet is divided into the $C$ and $H$ domains,
the $(6g-6)$-dimensional moduli space of 
${\cal M}_g$ becomes locally a product of the moduli spaces of 
$U_C$ and $U_H$. The simplest cases are when ${\cal S}=U_C$
or $U_H$, namely when the entire Riemann surface is
either in the $C$ domain or in the $H$ domain. 
The contribution from the case ${\rm \cal S}=U_C$
is denoted by ${\cal F}_{g,0}^{(C)}$, where $0$ refers to 
the fact that that is no boundary. 
We will compute this in (b) below. There is also a
possibility that the entire Riemann surface is in
the $H$ domain. This will be discussed in the next subsection.

Now let us consider the case when
 the boundary $\gamma$ has  one 
connected component, which separate the surface into
$U_C$ and $U_H$.
Suppose $U_C$ has $g_C$ handles and $U_H$ has $g_H$ handles
so that $g=g_C+g_H$. Among $(6g-6)$
Bertrami differentials on ${\cal S}$, we can choose 
$(6g_C-6)$ to be localized in $U_C$ and
$(6g_H-6)$ to be localized in $U_H$.
The remaining six moduli parameter are associated to
the long cylindrical region connecting $U_C$ and $U_H$
discussed in the last subsection. Among them, two
are associated to the location of an end point 
of the cylinder in $U_C$,
two are associated to the other end point of the
cylinder in $U_H$, 
and the remaining two describe the length $L$ and
the twist of the cylinder. If the surface ${\cal S}$
is separated into the two domains $U_C$ and $U_H$,
the twist parameter becomes irrelevant since we can
twist the cylinder along the boundary without
causing any change to $U_C$ or $U_H$. This is compensated
by the fact that the location $y_0$ of the boundary
$\gamma$ becomes an additional moduli parameter, so
the total number of moduli remains the same.   
We find $(6g_C-3)$ moduli parameters for $U_C$
and $(6g_H-3)$ moduli parameters for $U_H$. 

One can also see that the insertions of $G_L^-$
and $G_R^-$ in \gloop\ are distributed in the $C$
and $H$ domains in such a way that we obtain the correct measures
for the topological string amplitudes on $U_C$ and
$U_H$ with the boundary $\gamma$. This is obvious for
the moduli whose Beltrami differentials are localized 
in either $U_C$ or $U_H$.
The only slightly nontrivial ones are the insertions of $G_L^-$
and $G_R^-$ associated to the length and twist of the cylinder. 
These operators
are integrated along the $x$ direction 
as $\int_0^{2\pi} dx G_L^-(x,y) \ \int_0^{2\pi}
dx G_R^-(x,y)$ at some fixed $y$. We can move one of them 
to the $H$
domain $y < y_0$ and the other one to the $C$ domain
$y > y_0$, and use them to give the correct measures 
for the extra moduli.

Let us denote the amplitudes on $U_C$ and $U_H$ separated
by the single boundary $\gamma$ by 
where ${\cal F}_{g_C,1}^{(C)}$ and ${\cal F}_{g_H,1}^{(H)}$.
Taking into account the Jacobian factor \jacobiantwo ,
their contribution to the topological string amplitude
is
\eqn\singleboundary{
\left( {\cal F}_g \right)_{\# \gamma = 1} =
\sum_{g_C+g_H=g}
\Delta {\cal F}_{g_C,1}^{(C)} \cdot {\cal F}_{g_H, 1}^{(H)},}
where
\eqn\monodromy{
\Delta {\cal F}_{g_C,1}^{(C)} =
\oint d\sigma_0 {\partial {\cal F}_{g_C,1}^{(C)}\over \partial \sigma_0} 
}
computes the monodromy of ${\cal F}_{g_C,1}^{(C)}$
around $\sigma_0=0$.  This comes from doing
path integral over $\gamma$ leading to \gian\ as discuused before.
 Clearly the contribution vanishes
if ${\cal F}_{g_C,1}^{(C)}$ is a 
single-valued function of $\sigma_0$. This is the
case for $g_C \geq 1$ since the amplitude has a well-defined
expression as a correlation function of $(6g_C-3)$ supercurrents
$G^-$ integrated over the moduli space of surface with $g_C$
handles and one hole. The only case when the amplitude could potentially
have
a nontrivial monodromy around $\sigma_0=0$ is when $g_C=0$,
$i.e.$ when the $C$ domain has the topology of the disk. Note
that $6g_C-3=-3$ in this case, and the amplitude ${\cal F}_{0,1}^{(C)}$
itself is not defined as a correlation function of the A-model.
To compute it, we consider a well-defined correlation function
such  as $\partial^2{\cal F}_{0,1}^{(C)}/\partial \sigma_0 \partial t$
and integrate it with respect to $t$ and $\sigma_0$.
Actually
 it turns out  that
$\partial {\cal F}_{0,1}^{(C)}/\partial \sigma_0 $ is already
well defined, and we use it to compute $\Delta {\cal F}_{0,1}^{(C)}$.
We will show later that the disk amplitude
${\cal F}_{0,1}^{(C)}$ indeed has
a non-trivial monodromy around $\sigma_0=0$.
To summarize we have shown that, when the Riemann surface
is separated into one $C$ domain and one $H$ domain,
the only contribution to the topological string amplitude
comes from the case when the $C$ domain has a topology
of disk, as shown in Figure 8. 

\bigskip
\bigskip

\centerline{\epsfxsize 2.6truein \epsfysize 1.7truein\epsfbox{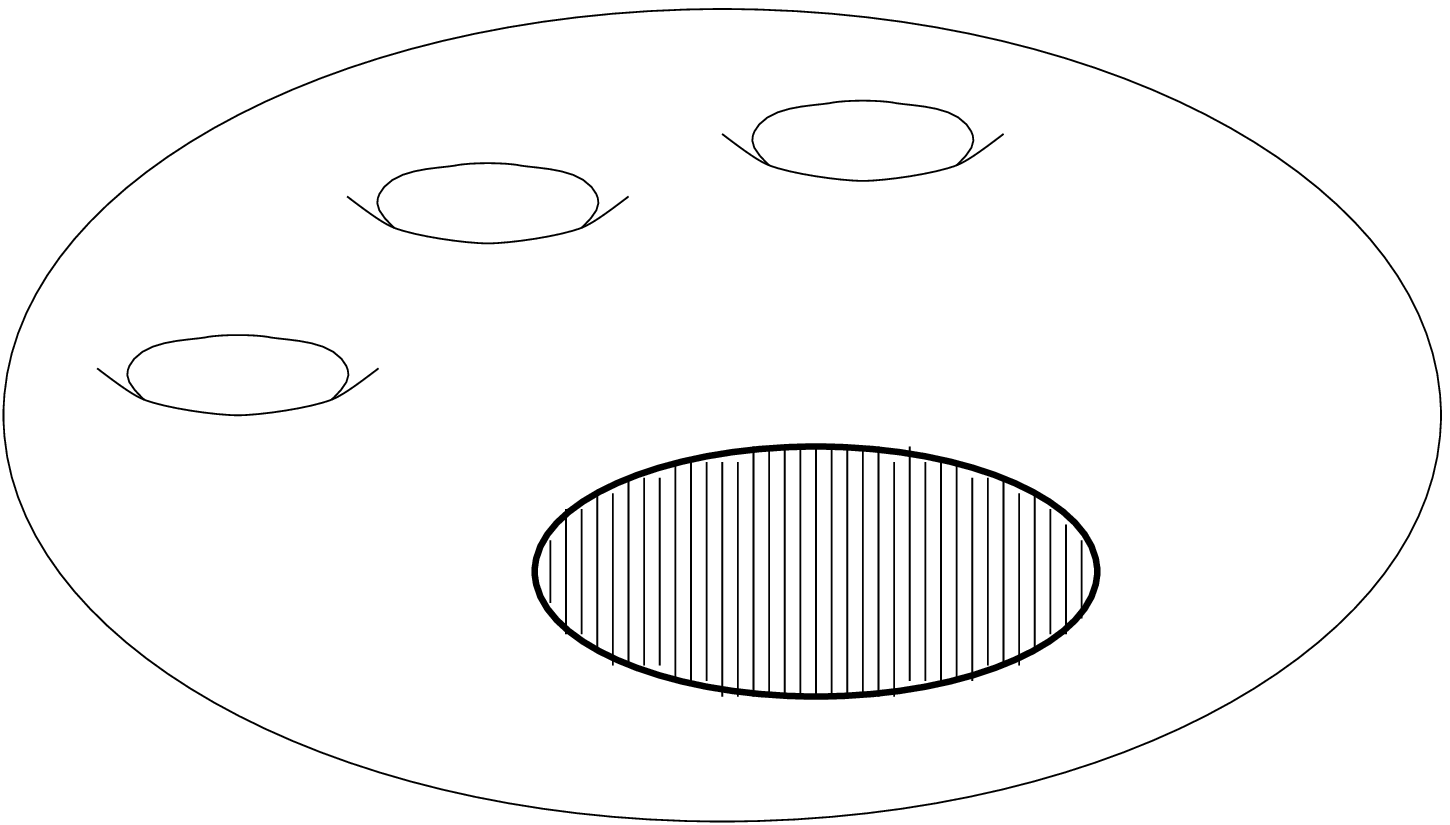}}
\noindent{\ninepoint\sl \baselineskip=8pt {\bf Figure 8}: {\sl
A worldsheet with both $H$ and $C$ domains contributes to the
topological string amplitude 
only when every $C$ domain has the topology of the disk.}}

\bigskip
\bigskip

It is straightforward to generalize this consideration to
the case when $\gamma$ has several connected components. 
The Jacobian factor \gian\ appears on each boundary
component.
We then perform the angular integral $\oint d\sigma_0$
for each boundary. We can show that 
the amplitude is non-zero
only when every connected component of $U_C$ has topology
of disk. Suppose $U_C$ has a connected component with
$g$ handles and $h$ boundaries. The topological string
amplitude computed on that component with the Dirichlet
boundary condition is well-defined if $6g+3h-6 \geq 1$.
Therefore, if $h\geq 1$, there are only two cases when 
the amplitude can have monodromy; $h=1$ or $2$ with
$g=0$, $i.e.$ it is either a disk or an annulus. 
Since the annulus has two boundaries, its contribution
to the topological string amplitude is
\eqn\annulus{ \oint d\sigma_0 \oint d\sigma'_0
{\partial^2 {\cal F}_{0,2}^{(C)}
\over \partial \sigma_0 \partial \sigma'_0}
,}
where ${\cal F}_{0,2}^{(C)}$ is the annulus amplitude
with the boundary conditions $\sigma=\sigma_0$ on
one boundary and $\sigma=\sigma'_0$ on the other.
Although ${\cal F}_{0,2}^{(C)}$ may have a monodromy,
its derivative $\partial {\cal F}_{0,2}^{(C)}/\partial \sigma_0'$
is  a single-valued holomorphic function
of $\sigma_0$. This means that the result of the $\sigma_0$ integral
for fixed $\sigma_0'$ in \annulus\ gives zero.
Therefore the mixed phase with the $C$ branch with the topology
of the annulus, as shown in Figure 9, does
not contribute to the topological string amplitude.
We have found that the $C$ domain must be a disjoint union of disks.

\bigskip
\bigskip

\centerline{\epsfxsize 2.8truein \epsfysize 1.7truein\epsfbox{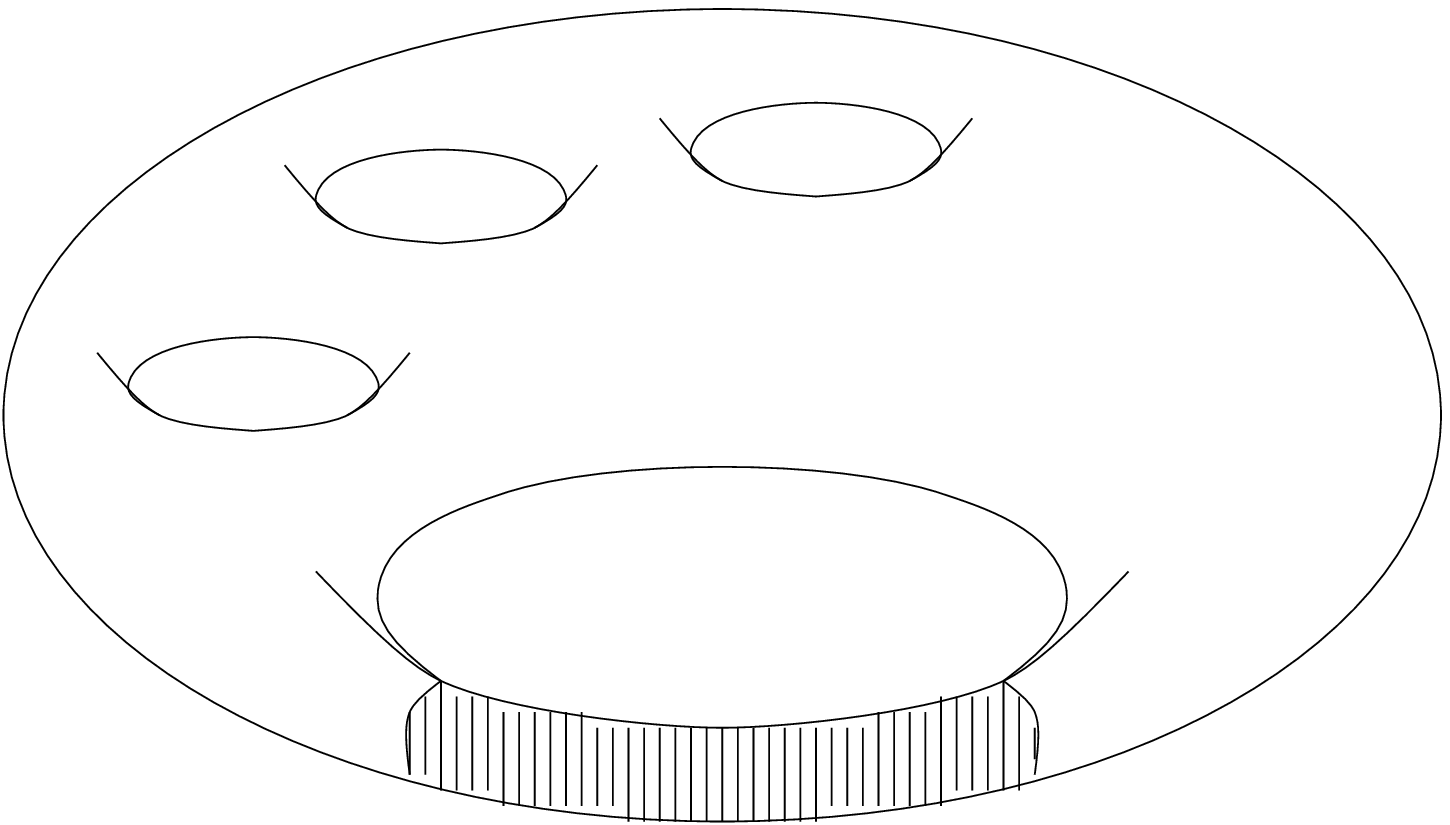}}
\noindent{\ninepoint\sl \baselineskip=8pt {\bf Figure 9}: {\sl
The partition function of the $C$ domain with the topology of
the annulus can have
a non-zero monodromy as a function of $\sigma_0$ and $\sigma_0'$,
which specify the Dirichlet boundary conditions on $\sigma(z)$.
However the amplitude is annihilated by
$\oint d\sigma_0 \oint d\sigma_0'
\partial_{\sigma_0} \partial_{\sigma_0'}$ and
therefore does not contribute to the topological string
amplitude.}}

\bigskip
\bigskip

We conclude the $g$ loop topological string amplitude
is expressed as
\eqn\sumoverholes{
{\cal F}_g = {\cal F}_{g,0}^{(C)} +
\sum_{h=1}^\infty \left(\Delta {\cal F}_{0,1}^{(C)}\right)^h
\cdot {\cal F}_{g, h}^{(H)} ,}
where $\Delta {\cal F}_{0,1}^{(C)}$ is the monodromy of the
disk amplitude and 
${\cal F}_{g,h}^{(H)}$ is the contribution from the $H$ domain
with $g$ handles and $h$ holes.

\medskip

Let us compute ${\cal F}_{0,1}^{(C)}$ and ${\cal F}_{g,0}^{(C)}$.

\bigskip
\noindent
(a) ${\cal F}_{0,1}^{(C)}$

\bigskip

As discussed above, in order to evaluate the contribution
of the disk amplitude we need to compute
$$\Delta {\cal F}_{0,1}^{(C)}
=\oint d\sigma_0 {\partial {\cal F}_{0,1}^{(C)}
\over \partial \sigma_0}.$$
The amplitude ${\cal F}_{0,1}^{(C)}$ is evaluated
in the $C$ domain of disk topology with the Dirichlet
boundary condition $\sigma=\sigma_0$. 
As noted in \lref\agv{M.~Aganagic and C.~Vafa,
``Mirror symmetry, D branes and counting holomorphic discs,''
{\tt arXiv:hep-th/0012041}.
}
\lref\kach{S. Kachru, S. Katz, A. Lawrence and
J. McGreevy, ``Open string instantons and superpotentials,''
Phys. Rev. D{\bf 62} 026001 (2000); {\tt arXiv:hep-th/9912151}.}
\lref\akv{
M.~Aganagic, A.~Klemm and C.~Vafa,
``Disk instantons, mirror symmetry and the duality web,''
Z.\ Naturforsch.\ A {\bf 57}, 1 (2002);
{\tt arXiv:hep-th/0105045}.
} \refs{\kach, \agv, \akv},
${\cal F}_{0,1}^{(C)}$ is given by the  integral
of the top holomorphic form bounding the brane. (In the case
of Calabi-Yau threefold this corresponds to a 2-cycle in Calabi-Yau
with the holomoprhic 3-form integrated over a 3-chain with boundary
being the 2-cycle.)
In the $C$ domain, we effectively
have a Landau-Ginzburg theory with superpotential
$$W=t\Sigma $$
$\Sigma$ is the twisted chiral superfield with $\sigma$
as its lowest component.  As discussed in \ref\hiv{K.~Hori, 
A.~Iqbal and C.~Vafa,
``D branes and mirror symmetry,''
{\tt arXiv:hep-th/0005247}.
}
for an ${\cal N}=2$ Landau-Ginzburg model, the top holomorphic
form is given by
$$\prod_i d\Phi_i e^{W(\Phi_i)}$$

In the case at hand, we have a single chiral field $\sigma$.  However
the measure in the field space is not $d\sigma$ but rather
\eqn\mes{{d\sigma \over \sigma^2},}
as noted in \siw
.  This arises
by integrating over the chiral fields $A_i,B_i$ which are massive
in the $C$ domain, which leads to a dilaton which translates to the above
change in measure. In the $C$ domain, the brane corresponds
to the 0-cycle specified by the point $\sigma =\sigma_0$
and thus the disk amplitude is an integral over a chain with
$\sigma_0$ as its boundary, $i.e.$
$${\cal F}_{0,1}^{(C)}(\sigma_0)
=\int^{\sigma_0} {d\sigma \over \sigma^2} e^{t\sigma}$$
Thus we have
$$\Delta {\cal F}_{0,1}^{(C)}
=\oint d\sigma_0 {\partial {\cal F}_{0,1}^{(C)}\over \partial \sigma_0}
=\oint d\sigma_0 {e^{t\sigma_0}\over \sigma_0^2}
\propto t,$$
which is what we wanted to show.  Thus each disk in the $C$ domain
contributes $t=N\lambda_s$ to the $H$ branch amplitude. This is
exactly what we need in order for the string amplitude to reproduce
the 't Hooft expansion \fg . 

\bigskip
\noindent
(b) ${\cal F}_{g,0}^{(C)}$

\bigskip

We now evaluate the amplitude when the entire Riemann surface
is in the $C$ domain.
There is an equivalent description of the $C$ domain in terms
of another Landau-Ginzburg theory, which can be motivated by trying
to make the measure in \mes\ be a more standard one.  If we define
the chiral field $X$ by
$$X=\Sigma^{-1},$$
 the superpotential is expressed as
\eqn\inversepot{W=t X^{-1},}
and the field measure changes to
$$dx=-{d\sigma\over \sigma^2}$$
Thus, in terms of $X$, the linear dilaton is turned off
and we have an ordinary Landau-Ginzburg theory. In fact this 
model has been proposed in \ref\mv{S.~Mukhi and C.~Vafa,
``Two-dimensional black hole as a topological coset 
model of $c = 1$ string theory,'' 
Nucl.\ Phys.\ B {\bf 407}, 667 (1993);
{\tt arXiv:hep-th/9301083}.
} as being related to the $c=1$
bosonic string with a target circle at self-dual radius.  
This follows from the fact that the Landau-Ginzburg model
with the superpotential \inversepot\
is equivalent to the $SL(2,R)/U(1)$ coset
conformal field theory at level $3$, which in turn
is related to the $c=1$ bosonic string. Originally
this equivalence was motivated by an ``analytic contiuation''
to $k=-3$  of the well-established
equivalence between the Landau-Ginzburg mode with
the superpotential $W=X^{k+2}$ and the minimal ${\cal N}=2$ conformal
field theory, which can be realized as the $SU(2)/U(1)$ 
coset model at level $k$. 
More recently, this equivalence was proven directly  
using mirror symmetry  in \ref\kah{
K.~Hori and A.~Kapustin,
``Duality of the fermionic $2d$ black hole 
and ${\cal N} = 2$ Liouville theory as  mirror symmetry,''
JHEP {\bf 0108}, 045 (2001);
{\tt arXiv:hep-th/0104202}.
}. 
In particular, the topological B-model
with the superpotential \inversepot\ is
 mapped to the topological A-model on $SL(2,R)/U(1)$
at level $3$.  In the $SL(2,R)/U(1)$ coset model, 
the topological string amplitude ${\cal F}_g(SL(2,R)/U(1))$
at genus $g$ has been 
computed \ref\witeu{E.~Witten,
``On the structure of the topological phase of two-dimensional gravity,''
Nucl.\ Phys.\ B {\bf 340}, 281 (1990).
}
with the result,
$${\cal F}_g(SL(2,R)/U(1))= {\chi \left({\cal M}_g\right)
\over  t^{2g-2}},$$
where $\chi\left( {\cal M}_g\right)$ is the Euler characteristic
of ${\cal M}_g$. 
More explicitly, 
\eqn\slpartition{ {\cal F}_g(SL(2,R)/U(1))
={B_{2g}\over 2g(2g-2)t^{2g-2}}}
for $g \geq 2$, where $B_{2g}$ is the $2g$-th Bernoulli number.
This can also be interpreted as the partition function of the $c=1$
bosonic string on a self-dual circle \ref\div{J.~Distler and C.~Vafa,
``A critical matrix model at $c = 1$,''
Mod.\ Phys.\ Lett.\ A {\bf 6}, 259 (1991).
}. From the above chain of equivalences, we find that  
\slpartition\ gives the contribution to the topological string
amplitude when the whole Riemann surface is in the $C$ domain.

To conclude, we have
\eqn\cdomain{ {\cal F}_{g,0}^{(C)} = {B_{2g} \over 2g(2g-2) t^{2g-2}},} 
for $g \geq 2$.
The topological string amplitudes at $g=0$ and $1$ are somewhat special.
For $g=0$, we have
\eqn\cdomainone{{\cal F}_{0,0}^{(C)}={1\over 2}t^2 {\rm log}t+Q(t),}
where $Q(t)$ is a quadratic polynomial in $t$.  The easiest
way to see that is to note that 
$\partial{\cal F}_{0,0}^{(C)}/\partial t$ 
is a period integral
given by 
\eqn\cdomaintwo{\partial{\cal F}_{0,0}^{(C)}/\partial t=
\int d\sigma \sigma^{-2} e^{t\sigma}.}
Taking two more derivatives with respect to $t$ allows one
to do the integral and one obtains 
$\partial^3{\cal F}_{0,0}^{(C)}/\partial t^3=t^{-1}$ 
which can be integrated to find the above result.  The genus $1$
amplitude is 
\eqn\cdomainthree{{\cal F}_{1,0}^{(C)}=-{1\over 12}
\log t.}
We find that \cdomain , \cdomainone\ and \cdomainthree\
reproduce the nonperturbative terms \annon\ 
in the Chern-Simons theory amplitude that are not
captured in the 't Hooft expansion.

As an aside, we would like to point out that 
the above results also provide another insight into the
claim that the Landau-Ginzburg B-model with $W=tX^{-1}$
is equivalent to the B-model
conformal theory on the {\it deformed} conifold \ghv . 
This follows from the fact that the most
singular contribution to A-model amplitudes
on {\it resolved} conifold when $t\rightarrow 0$
should come from the $C$ domain which opens up at that point, 
giving rise to the results in the above paragraphs.  
On the other hand in this limit, the mirror symmetry
maps the resolved conifold to the deformed conifold.
This explains the equivalence between the B-model
on the deformed conifold and the above Landau-Ginzburg B-model.  

Thus we have a number of equivalences: 

\smallskip
\noindent
$\bullet$ The topological Landau-Ginzburg B-model with
the superpotential $W=t\Sigma$ with the non-standard
measure $d\sigma/\sigma^2$. 

\smallskip
\noindent
$\bullet$ The topological Landau-Ginzburg B-model with
the superpotential $W=tX^{-1}$ with the standard measure.

\smallskip
\noindent
$\bullet$ The $t\rightarrow 0$ limit of the topological A-model
on the resolved conifold. 

\smallskip
\noindent
$\bullet$ The $SL(2,R)/U(1)$ coset A-model.

\smallskip
\noindent
$\bullet$ The $c=1$ string theory at the self-dual radius.

\smallskip
\noindent
Moreover the contribution from this piece 
is precisely the part that the perturbative Chern-Simons
theory lacks from the exact results.

Note that these equivalences also gives rise to an {\it a priori}
formula relating the volume of $U(N)$ to the Euler characteristic
$\chi({\cal M}_g)$ of the moduli space of genus $g$ Riemann
surfaces as 
\eqn\interesting{ {\rm log}\left({\rm vol}\left(U(N)\right)\right)
=-\sum_g {\chi({\cal M}_g) \over N^{2g-2}} , }
with an appropriate interpretion for $g=0$ and $1$ terms. 
To our knowledge, this intriguing 
formula has not been noted in any 
matrix model or topological string literature.

\subsec{Contribution from the $H$ domain}

In the $H$ domain, the functional integral over $\sigma$
is cutoff at $|\sigma|<\sigma_*$ and therefore is regular
in the conifold limit. We can regard this domain as describing
topological string propagating in the smooth part of
the conifold. As explained in section 2, the smooth resolution
is a local operation at the conifold singularity, so we expect
that amplitudes computed in the $H$ domain to be independent
of the K\"ahler moduli $t$. In fact, since the $U(1)$ gauge
symmetry is spontaneously broken in the $H$ domain,
the field strength $F$ is zero and there is no contribution
from the theta term, $i {\theta \over 2\pi} F$. By holomorphy
in $t$, we conclude that the amplitude in the $H$ domian
does not depend on $t$. Thus we will evaluate the contribution
from the $H$ domain right at the conifold limit $t=0$. 

In the $C$ domain, the chiral multilet fields $a_i, b_i$ are
suppressed due to the potential \potential\ with $|\sigma| > \sigma_*$.
Viewed from the $H$ domain, we have the Dirichlet condition
$a_i, b_i=0$ at the boundary. Namely the chiral multiplet fields
in the $H$ domain is ending on the conifold singularity.
To evaluate the amplitude, it is useful to regard
the conifold as the $\mu \rightarrow 0$ limit of 
the deformed conifold geometry \deformedconi .
Since the deformation parameter $\mu$ is BRST trivial
in the topological A-model, the amplitude is independent
of $\mu$. Thus we can turn on $\mu$ without changing 
the amplitude. 
To do this, we need to
use $y_i$ variables which are related to $a_i, b_i$ by 
\generalsol\ when $\mu=0$. 
When $\mu >0$, the deformed conifold is the cotangent
bundle over $S^3$, as we saw in section 2, and the natural
boundary condition for the A-model is that the open
string ends on the base $S^3$. In the phase space
coordinate \phasecoord, the conditions are
$p_i=0$, which correspond in the $y_i$ coordinates to
\eqn\lagrangian{ y_1= \bar y_2, ~~y_3 = -\bar y_4.}
In the limit $\mu \rightarrow 0$, these conditions together with
the equation for the conifold geometry \conitwo\ 
imply $y_i=0$ and reduces to the description in terms
of $a_i, b_i$ with the fully Dirichlet condition on them. 
Since the A-model amplitude is independent of $\mu$, 
${\cal F}_{g,h}^{(H)}$ should be given by the topological
string amplitude with D brane wrapping on the base $S^3$
of $T^* S^3$. According to \wittenjones , this is
nothing but the sum over Feynman diagrams of the Chern-Simons
gauge theory on $S^3$ denoted by $F_{g,h}$ in section 3.
Here
we are obtaining the strictly zero size limit of $S^3$.  However
when we deform $S^3$ to finite size, the topological
A-model amplitudes do not change, and the only allowed D-brane
in the geometry is the one corresponding to the branes wrapping $S^3$.
Thus, smoothing out the geometry gives the uniquely allowed branes,
yielding the description of $U(N)$ Chern-Simons on $S^3$.

As we mentioned in the last subsection, there is also
the possibility that the entire Riemann surface is in
the $H$ domain. This would describe the closed topological
string on the conifold with finite size $S^3$, 
and its amplitudes is trivial 
since there is no two-cycles which is needed for
worldsheet instantons.\foot{Even degenerate instantons
do not contribute in this case.} This is consistent with
the expectation that the closed string degrees of freedom
should have been decoupled in the gauge theory side.  

\subsec{Summary}

We have found that, by summing over configurations
of the $C$ and $H$ domains, the $g$ loop topological
string amplitude for $g \geq  2$ is given by
\eqn\final{{\cal F}_g=
{B_{2g}\over 2g(2g-2)t^{2g-2}} \ + \
 \sum_{h=1}^\infty \ t^h  F_{g, h}.}
The sum over $h$ in the second term reproduces
the 't Hooft expansion of the Chern-Simons gauge theory, 
and the coefficients $F_{g, h}$ are computed using 
ribbon graphs with
$g$ handles and $h$ holes. On the other hand, the first term
comes from the situation when the entire Riemann surface
is in the $C$ domain. This term is missing in the
't Hooft expansion, as discussed in section 2.

The formulae for $g=0$ and $1$
are given by
\eqn\finaltwo{\eqalign{
{\cal F}_{g=0} &= {t^2\over 2} \log t  +  Q(t) + 
  \sum_{h=1}^\infty \ t^h
 F_{g=0, h}, \cr
{\cal F}_{g=1} &= -{1\over 12} \log t
+  \sum_{h=1}^\infty \ t^h
 F_{g=1, h},}}
where $Q(t)$ is some quadratic polynomial of $t$.
Combining \final\ and \finaltwo\ together and
coverting the variables as $t = g_{{\rm YM}}^2 N$ and
$\lambda_s = g_{{\rm YM}}^2$, the topological
string amplitude to all order in the perturbative
expansion is given by
\eqn\allgenera{
\eqalign{ {\cal F} =&
\sum_{g=0}^\infty g_{{\rm YM}}^{2g-2} {\cal F}_g^{(C)} \cr
=& {N^2 \over 2} \log \left( g_{{\rm YM}}^2 N \right) 
-{1\over 12} \log \left( g_{{\rm YM}}^2 N \right)
+ \sum_{g=2}^\infty {1 \over N^{2g-2}}
{B_{2g}\over 2g(2g-2)} \cr
& ~~~~~~~~~+ g_{{\rm YM}}^{-4}
Q(g_{{\rm YM}}^2 N) \cr
&+ \sum_{g=0}^\infty \sum_{h=1}^\infty (g_{{\rm YM}})^{2g+h-2}
N^h F_{g, h}.}}
The sum over $g$ and $h$ in the last term in the right-hand
side reproduces the 't Hooft expansion of the Chern-Simons
gauge theory. 
The non-perturbative conributions
of the gauge theory are also in perfect agreement
with the terms in the above coming from the case when the entire
Riemann surface is in the Coulomb branch.

\subsec{$SO(N)$ and $Sp(N)$ gauge groups}

It was conjectured in 
\ref\SinhaAP{
S.~Sinha and C.~Vafa,
``$SO$ and $Sp$ Chern-Simons at large $N$,''
{\tt arXiv:hep-th/0012136}.
}
that the Chern-Simons gauge theories on $S^3$ with
$SO(N)$ and $Sp(N)$ gauge groups are dual to 
the topological closed string theories 
on an orientifold of the resolved
conifold, with the dictionary that 
\eqn\dictionary{ \eqalign{\lambda_s & = {i \over k+c_g}, \cr
              t & = (N+a) \lambda_s,}}
with $c_g=N-2, a = -1$ for $SO(N)$ and
$c_g=N + 1, a = +1$ for $Sp(N)$.
It should be possible to extend the proof 
in this paper to these cases with the orientifolds.

In particular, 
this means that the identity \interesting\
for $U(N)$ can be generalized and that there is a relation 
between the volumes of $SO(N)$ and $Sp(N)$ gauge groups and virtual
Euler characteristics on the moduli spaces of orientable and non-orientable
Riemann surfaces. 
The volumes of these groups can be
evaluated using the formulae in \lref\hashimoto{Y. Hashimoto,
{\it unpublished notes}.} \refs{\macdonald, \hashimoto}  
\eqn\othergroups{
\eqalign{
{\rm vol}(SO(2n+1)) & =
{2^{n+1} (2\pi)^{n^2+n-{1\over 4}}
 \over (2n-1)! (2n-3)! \cdots 3! 1!} , \cr
{\rm vol}(SO(2n)) & = {\sqrt{2} (2\pi)^{n^2}
\over (2n-3)! (2n-5)! \cdots 3! 1!\ (n-1)! } , \cr
{\rm vol}(Sp(2n)) & =
 {2^{-n} (2\pi)^{n^2+n}
\over (2n-1)! (2n-3)! \cdots 3! 1!}.}}
The factorials in the denominator of each formula
is related to the exponents of the corresponding group.
Note that the expressions for ${\rm vol}(SO(2n))$ and
${\rm vol}(SO(2n+1))$ are related to each other by
analytic continuation in $n$. This can be shown
by using 
\eqn\shiftformula{ \eqalign{ (2n-3)! (2n-5)! \cdots 3! 1!\ (n-1)! 
\ \rightarrow& \ 2^{-n}\sqrt{\pi} (2n-1)! (2n-5)! \cdots 3! 1!\cr
& \left(n \rightarrow n + {1\over 2}\right).}}
The analyticity in $n$ 
is necessary in order for the large
$n$ expansion to make sense. 
One can also check explicitly that these formulae
agree with the $k \rightarrow \infty$ limit of
the exact results of the $SO$ and $Sp$ 
Chern-Simons gauge theory. 

Let us evaluate
the large $n$ expansion of the logarithm of each
of these volume formulae and compare with 
the virtual Euler characteristics of the moduli spaces.
As pointed out in \SinhaAP , the non-orientable Riemann surfaces
with two cross caps do not contribute to the topological closed
string amplitudes.  Therefore topological amplitudes with
the orientifold of the $SO$ ($Sp$) types
are equal to one-half of the amplitude without
the orientifold plus (minus) contributions with one
crosscap. This means that the product of the
volume of the $SO$ and $Sp$ groups should be
related to the volume of the unitary group.  
To be precise, we need to take into account the
shift in $N \rightarrow N+a$ in the expression 
for $t$ in \dictionary . 
Using the volume formulae \othergroups\ and
performing the analytic continuation $2n \rightarrow 
2n - a$ to compensate for the shift, we find
\eqn\productvolume{
\eqalign{&
 {\rm vol}\left(SO(2n+1)\right) \cdot {\rm vol}
\left( Sp(2n-1) \right) \cr
& =\sqrt{4\over \pi}
{(2\pi)^{2n^2 + n} \over
  (2n-1)! (2n-2)! \cdots 2! 1!}.}}
Comparing this with \volumeun 
, we find
\eqn\comparison{
 {\rm vol}\left(SO(2n+1)\right) \cdot {\rm vol}
\left( Sp(2n-1) \right)
= \sqrt{4\over \pi}
 {\rm vol}\left( U(2n)\right).}
Thus, using \interesting ,  we can write 
\eqn\moreinteresting{\eqalign{&
\log \left( {\rm vol}\left( SO(2n+1)\right) \right)
 + \log \left( {\rm vol}\left( Sp(2n-1)\right) \right) \cr
& = - \sum_g {\chi({\cal M}_g)\over (2n)^{2g-2}}.}}
As in the case of $U(N)$, the $g=0$ and $1$ terms
need to be interpreted appropriately.

Similarly the large $N$ duality predicts that the
logarithm of the ratio of these volumes gives a
generating function of 
the virtual Euler characteristic 
$\chi({\cal M}^1_g)$ is the Euler characteristic
of moduli space of genus $g$ with a single cross cap as,
\eqn\mreinteresting{\eqalign{&
\log \left( {\rm vol}\left( SO(2n+1)\right) \right)
 - \log \left( {\rm vol}\left( Sp(2n-1)\right) \right) \cr
& =\log\left({ 2^{-{1\over 2}} (4\pi)^{n+{1\over 2}}\over 
\Gamma(n+{1\over 2})}\right)=  
- \sum_g {\chi({\cal M}^1_g)\over (2n)^{2g-1}}.}}
Using the expansion of the Gamma function, we find
$$ \chi({\cal M}^1_g)={(2^{2g-2}-2^{-1})B_{2g}\over 2g(2g-1)}.$$
This is in full agreement with the recent computation 
of this quantity \ref\haret{I.P.Goulden, J.L. Harer, and D.M. Jackson,
``A geometric parametrization for the virtual Euler characteristic
of the moduli spaces of real and complex algebraic curves,''
Trans. Am. Math. Soc. {\bf 353}, 4405 (2001).}.

To summarize, we found
\eqn\soandsp{
\eqalign{
\log\left( {\rm vol}\left(SO(2n+1)\right)\right)
& = -{1\over 2} \sum_g \left( {\chi({\cal M}_g)\over (2n)^{2g-2}} +
{\chi({\cal M}_g^1) \over (2n)^{2g-1}}\right),\cr
\log \left( {\rm vol}\left(Sp(2n-1)\right)\right)
& = -{1\over 2} \sum_g \left( {\chi({\cal M}_g)\over (2n)^{2g-2}} -
{\chi({\cal M}_g^1)\over  (2n)^{2g-1}}\right).}}
As predicted by the large $N$ duality,
the volume factors in the $SO$ and $Sp$ Chern-Simons
gauge theory are related to the topological
string computation for orientable and non-orientable 
worldsheets in the pure $C$ domain. 

\newsec{Discussion}

We have proven that the topological closed string
amplitude ${\cal F}$ on the resolved conifold geometry can be 
expressed as
\eqn\conclusion{
{\cal F} = {\cal F}_{{\rm nonpert}} +
 \sum_{g=0}^\infty \sum_{h=1}^\infty (g_{{\rm YM}})^{2g+h-2}
N^h F_{g, h},}
where $F_{g,h}$ is the Chern-Simons amplitudes on $S^3$
computed by using ribbon graphs with $g$ handles and $h$ holes.
In the closed string theory, $F_{g,h}$ computed using genus $g$
Riemann surfaces on which there are $h$ disks in the $C$ phase. 
Seen from the $H$ phase, 
these disks in the $C$ phase are holes on the worldsheet,
which introduce an open string sector. We showed that,
when the $C$ and $H$ phases coexist, each $C$ domain
must have the topology of the disk. 
Moreover we found that 
each disk in the $C$ domain 
gives the factor $t = g_{YM}^2 N$, precisely as 
prescribed for the 't Hooft expansion. 
 In this way, the closed string computation reproduces
the Chern-Simons perturbative expansion.

The extra term ${\cal F}_{{\rm nonpert}}$ 
is given by a sum over Riemann surfaces
which are in the pure $C$ phase. 
This term, which cannot be captured
in the 't Hooft expansion of the gauge theory,
 has various
interpretations. 
From the string theory point of view, they
are:

\medskip

\noindent
$\bullet$ The topological string coupled to the
topological Landau-Ginzburg
model with the superpotential $W=tX^{-1}$.

\smallskip
\noindent
$\bullet$ The topological string coupled to the
$SL(2,R)/U(1)$ coset A-model.

\smallskip
\noindent
$\bullet$ The $c=1$ topological string amplitude on a circle
of self-dual radius.

\medskip
\noindent
These compute the Euler characteristic 
$\chi({\cal M}_g)$ of the moduli space of genus $g$ Riemann
surfaces. From the gauge theory point of view, they are:

\medskip
\noindent
$\bullet$ The measure factor of the Chern-Simons
gauge theory on $S^3$.

\smallskip
\noindent
$\bullet$ The fractional instanton contribution
to superpotential terms on D6 branes
wrapping on the $S^3$ of the deformed conifold geometry.

\medskip
\noindent
These compute the volume of $U(N)$. This leads to the equality
\interesting\ relating $\chi({\cal M}_g)$ to the volume of $U(N)$. 
All these pictures fit together to make the gauge theory/string
theory equivalence complete. 

\bigskip

In this paper, we have presented the proof of the large $N$ 
duality using the partition function of the Chern-Simons theory
as an example. It is not difficult to apply our derivation
to other physical observables, such as expectation values of 
Wilson loops, which compute knot and link invariants.
The proof goes as follows. 
As shown in \oov , the computation involving Wilson loops
can be formulated in the language of topological
open string theory by adding probe D branes on 
$T^* S^3$ wrapping on special Lagrangian submanifolds 
which are determined by the configuration of the loops 
on $S^3$. Turning on holonomies of the gauge fields
on the probe branes, the partition function of the topological
open string theory gives a generating function of the
knot and link invariants.\foot{The generating function
is a function of the gauge field holonomies on the probe
branes and carries information about Wilson loops
expectation values for arbitrary representations of $U(N)$
running around the loops.}
 The large $N$ duality predicts 
that the generating function can be computed as 
a partition function of the topological string on 
the resolved conifold in the presence of 
the corresponding probe D branes. 
It was pointed out in \marv\ that, in order to fix the framing 
ambiguity of the loops, one needs to move the probe D branes 
off from the base $S^3$, adding mass to the open string stretched 
to the probe D branes and the ones wrapping on the $S^3$. 
It is then clear what the corresponding branes are in the string
theory side since the small resolution only modifies
the geometry near the conifold point. 

Given this, we can prove the duality involving the Wilson loops
as follows. We start from the topological closed
string on the resolved conifold with these probe D branes.
We note that only the $H$ domain of the worldsheet
can end on the probe D branes since they are away from
the conifold point. We can then run the same argument as
we presented in section 4 and show that this reproduces
the topological open string computation with the $N$
D branes wrapping on the base $S^3$ as well as the probe
branes. In this way, the partition function of the topological
closed string theory with the probe D branes can be shown
to reproduce the generating
function of the knots and links.
In this case, we do not get the extra contribution
from worldsheets in the pure $C$ phase since we need
the $D$ domain in order for the worldsheet to end
on the probe branes. 

\bigskip

The proof of the large $N$ duality was made possible by the
linear sigma-model description which is useful when the 
't Hooft coupling $t = g_{{\rm YM}}^2 N$ is small.   It is in some
sense a ``non-perturbative'' completion of the non-linear sigma model.
Namely, if we had considered strictly the non-linear sigma model, we
would have concluded that the worldsheet theory becomes singular 
at $t\rightarrow 0$. In the linear sigma model description, 
some massive states become light and relevant as new degrees
of freedom in the infrared limit.  
It is somewhat similar to how the non-perturbative
string states become light in the target physics, and are 
important in understanding
the target dynamics.
It would be interesting to come up with a similar description
for strings in $AdS$, which should enable us to prove
the $AdS$/CFT correspondence from the first principle.
We need an analogue of the linear sigma model for $AdS$
which gives a useful description of the worldsheet when 
the curvature radius of $AdS$ vanishes.

\bigskip
\bigskip

\centerline{{\bf Acknowledgments }}

\bigskip
We thank M. Aganagic, R. Dijkgraaf, 
M. Douglas, R. Gopakumar, Y. Hashimoto, S. Katz, M. Marino, A. Strominger,
N. Warner and E. Witten
for valuable discussions.
H.O. would like to thank the hospitality of
the theory group at Harvard
University. 
C.V. would like to thank the hospitality of the YITP at Stony
Brook.

The research of HO was supported in part by
DOE grant DE-FG03-92-ER40701. The research
of CV was supported in part by NSF grants
PHY-9802709 and DMS-0074329.

\vfill

\listrefs

\bye